\documentclass[10pt]{article}
\usepackage{mathrsfs}
\usepackage{amsmath,amsfonts,amssymb,mathtools}
\usepackage{cite}
\usepackage{dsfont}
\usepackage{graphicx}
\usepackage{hyperref}
\usepackage{verbatim}
\usepackage{slashed}
\usepackage[all]{xy}
\usepackage{xcolor}
\usepackage{subfig}
\usepackage{tikz}
\usetikzlibrary{shapes.geometric,arrows}
\usepackage{relsize}
\usepackage{caption}
\usepackage{float}
\usetikzlibrary{positioning}
\usepackage{geometry}
\usepackage[T1]{fontenc}
\usepackage{lmodern}
\usepackage{stackengine} 
 
\usepackage[utf8]{inputenc}

\hypersetup{
  colorlinks,
  citecolor=violet,
  linkcolor=blue,
  urlcolor=blue}

\csname @addtoreset\endcsname{equation}{section}
\DeclareMathAlphabet\mathbfcal{OMS}{cmsy}{b}{n}

\newcommand{\beq}{\begin{equation}}
\newcommand{\eeq}{\end{equation}}
\newcommand{\bea}{\begin{eqnarray}}
\newcommand{\eea}{\end{eqnarray}}
\newcommand{\ba}{\begin{array}}
\newcommand{\ea}{\end{array}}
\newcommand{\bit}{\begin{itemize}}
\newcommand{\eit}{\end{itemize}}
\newcommand{\nn}{\nonumber}

\newcommand{\mezzo}{\frac{1}{2}}
\newcommand{\complesso}{{\ \hbox{{\rm I}\kern-.6em\hbox{\bf C}}}}
\newcommand{\reale}{{\hbox{{\rm I}\kern-.2em\hbox{\rm R}}}}
\newcommand{\uno}{ \,  \raisebox{+0.14em}{{\hbox{{\rm \scriptsize ]}} \raisebox{-0.2em}{\kern-.8em\hbox{1}}}} \, }  

\newcommand{\p}{\partial}

\newcommand{\w}{\wedge}

\renewcommand{\a}{\alpha}

\newcommand{\g}{\gamma}

\newcommand{\D}{\Delta}
\newcommand{\e}{\epsilon}
\newcommand{\Er}{{\mathbfcal{E}}}

\renewcommand{\l}{\lambda}
\renewcommand{\L}{\Lambda}

\newcommand{\m}{\mu}

\newcommand{\n}{\nu}
\renewcommand{\r}{\rho}
\newcommand{\s}{\sigma}

\renewcommand{\S}{\Sigma}

\newcommand{\om}{\omega}
\newcommand{\Om}{\Omega}

\begin{document}


\begin{titlepage}

\begin{flushright}
$IFUM$--1105--$FT$ \\
$LIFT$--3-1.23
\end{flushright}

\vspace{0.3cm}

\begin{center}
\renewcommand{\thefootnote}{\fnsymbol{footnote}}
\vskip 10mm  
{\Huge \bf Plebanski-Demianski goes NUTs 
\vskip 6mm
  (to remove the Misner string)}
\vskip 26mm
{\large {Marco Astorino$^{a}$\footnote{marco.astorino@gmail.com},
Giovanni Boldi$^{b}$\footnote{giovanni.boldi@studenti.unimi.it}
}}\\

\renewcommand{\thefootnote}{\arabic{footnote}}
\setcounter{footnote}{0}
\vskip 8mm
\vspace{0.2 cm}
{\small \textit{$^{a}$Laboratorio Italiano di Fisica Teorica (LIFT),  \\
Via Archimede 20, I-20129 Milano, Italy}\\
} \vspace{0.2 cm}

{\small \textit{$^{b}$Dipartimento di Fisica, Universit\`a degli Studi di Milano,}} \\
{\small {\it Via Celoria 16, I-20133 Milano, Italy}\\
}
\end{center}

\vspace{2.3 cm}

\begin{center}
{\bf Abstract}
\end{center}
{We present a general procedure, based on the Ehlers transformation of the Ernst equations, to add the gravitomagnetic mass to the whole Plebanski-Demianski family of solutions. We can efficiently generate a large class of accelerating black holes, such as Reissner-Nordstrom or Kerr-Newman, endowed with the NUT parameter. The full rotating version carries a couple of independent NUT charges, one associated to the black hole and the other to the accelerating Rindler background. \\
The two NUT parameters can be coupled to remove the axial irregularity which causes the Misner string, still remaining with a Lorentzian spacetime, without the need to impose periodic time.\\
All the metrics we build are not of D-type according to the Petrov classification, but type-I, so they belong to a more general category with respect to C-metrics and the Plebanski-Demianski seed.\\
A convenient form of the most general type D black hole solution in general relativity, coupled with Maxwell electromagnetism, is obtained when switching off one of the two NUT parameters.}

\end{titlepage}

\addtocounter{page}{1}

\newpage


\section{Introduction}

Recently a renewed interest for accelerating and NUTty black holes has emerged, despite these spacetimes often bringing problematic physical features, such as cosmological or Misner strings, axial singularities or closed timelike curves. Some of these characteristic can be removed in the case of accelerating black holes without the Newman-Unti-Tamburino (NUT) parameter, as showed firstly by Ernst by means of external electromagnetic or gravitational fields \cite{ernst-remove}, \cite{ernst-generalized-c} (see also \cite{marcoa-pair} \cite{multipolar-acc} for generalizations) or by means of the gravitational spin-spin interaction \cite{marcoa-removal}. On the other hand it seems harder at the moment to fix the mathematical and physical criticalities of black holes which carry the NUT parameter, neither with standard time, which flows in an unbounded range, nor with periodical identification for the time coordinate, as first discussed by Misner in \cite{misner-counterexample}. \\ %
The Plebanski-Demianski (PD) family of solutions \cite{Plebanski-Demianski} is a good starting point to study such NUTty and accelerating black holes. Indeed it is well known it describes a large class of type D black holes in General Relativity. Namely the PD class comprises accelerating, electromagnetically charged, stationary rotating metrics. In certain subcases, for instance in absence of acceleration, these solutions are endowed also with an additional integrating constant related to the gravitomagnetic mass, sometimes also known in the literature as ``NUT charge''. On the other hand, when the acceleration parameter is switched on, the coexistence of both acceleration and gravitomagnetic mass seems to lie outside the Plebanski-Demianski family of solution and even the type D class of spacetimes. In fact, as shown in \cite{Podolsky-2020}, even the most simple case, constituted by the accelerating Taub-NUT metric, belongs to the general Type-I class of the Petrov classification. The only solution of this kind was found in \cite{mann-stelea-chng}, in a cumbersome way. That procedure is so involved that since then it was not possible to extend it to more general cases. Moreover the resulting metric was not easy to read and analyse until recently it has been cast in a convenient form in \cite{Podolsky-2020}. \\ %
The scope of this article is to find a method to generate this kind of solutions in an efficient way and in a suitable form, in order to be able to generalise the accelerating Taub-NUT metric in the presence of rotation and Maxwell electromagnetic charges. We plan to exploit one of the symmetries, the Ehlers transformation, of the Ernst equations. These equations model the axisymmetric and stationary spacetimes in the Einstein-Maxwell theory. In particular with this generating technique, we expect to build the accelerating Reissner-Nordstrom-NUT and the accelerating Kerr-(Newman)-NUT solutions, which are unknown spacetimes, in the literature, so far. This point will be developed in sections \ref{sec:acc-RN-Nut} and \ref{sec:generation} respectively.\\
It is commonly believed that the Plebanski-Demianski metrics already carry the NUT parameter, but in the presence of acceleration this fact is true only in the rotating case: when the angular momentum parameter vanish from the PD solution, also the acceleration disappears, so we cannot have a limit where both acceleration and NUT parameter coexist at the same time. On the other hand we will show how our procedure is able to add an extra independent NUT parameter either in the presence of angular momentum or not, still generating a non-type D black hole. Therefore when the accelerating seed already carries a NUT parameter we can remain with two different NUT charges. Note that this is not trivial because in absence of acceleration the NUT parameter, appended by the Ehlers transformation to a Taub-NUT or Kerr-NUT seeds, is not really independent, in fact it can be absorbed in the initial parameters of the seed, without changing the physical properties of the metric. Thus in these non accelerating cases the Ehlers transformation does not change the physical properties of the spacetime but it basically acts trivially, as the identity transformation.    \\
We begin, in section \ref{sec:Ernst-Ehlers}, with a brief introduction of the Ehlers transformation in general relativity coupled with the electromagnetic field, in the context of the Ernst reformulation of the Einstein-Maxwell equations.

\section{Ernst equations and Ehlers transformation}
\label{sec:Ernst-Ehlers}

In this article we will consider pure general relativity coupled with a standard electromagnetic field in four space-time dimensions, governed by the Einstein-Hilbert action
\beq
         I[g_{\m\n},A_\m] = - \frac{1}{16\pi G} \int_\mathcal{M} d^4x \sqrt{-g} \left(R - F_{\m\n}F^{\m\n} \right) \ . \nn
\eeq
From the above action we can derive the Einstein-Maxwell equations of motion
\bea  \label{field-eq-g}
                        &&   R_{\m\n} -   \frac{R}{2}  g_{\m\n} =   F_{\m\r}F_\n^{\ \r} - \frac{1}{4} g_{\m\n} F_{\r\s} F^{\r\s}  \quad ,   \\
       \label{field-eq-A}                  &&   \partial_\m ( \sqrt{-g} F^{\m\n}) = 0  \ \quad , 
 \eea
where the Faraday electromagnetic tensor stems from the vector potential $A_{\m}$ such that $F_{\m\n} = \p_\m A_\n -\p_\n A_\m$. 
Since we are interested only in integrable axisymmetric and stationary solutions we choose the gauge potential as follows
\beq \label{Am}
A_{\mu}=\big( A_t , 0 , 0 , A_\varphi \big) \ \ .
\eeq
Similarly the most general metric compatible with the above symmetries, i.e. two commuting Killing vector fields $(\p_t,\p_\varphi)$,  is the Lewis-Weyl-Papapetrou metric
\beq
\label{LWP-metric}
{ds}^2 = -f ( dt - \omega d\varphi)^2 + f^{-1} \bigl[ e^{2\gamma}  \bigl( {d \rho}^2 + {d z}^2 \bigr) +\rho^2 d\varphi^2 \bigr] \ \ .
\eeq
The three structure functions of the metric $f,\omega,\g$, the electric and magnetic potentials $A_t , A_\varphi$  depend only on the non-Killing coordinates, to preserve the requested symmetry hypothesis. It was shown in \cite{ernst2} that Einstein-Maxwell equations of motion (\ref{field-eq-g})-(\ref{field-eq-A}) are basically equivalent to the complex Ernst equations\footnote{The couple of equations necessary to determine $\g$ are not included, but they are completely decoupled from the other functions $f,\omega,A_t,A_\varphi$, therefore can be deduced by a couple of integrals in a subsequent step, for more details coherent with this notation see \cite{enhanced}.}
 \bea 
     \label{ee-ernst-ch}  \left( \textsf{Re} \ \Er + | \mathbf{\Phi} |^2 \right) \nabla^2 \Er   &=&   \left( \overrightarrow{\nabla} \Er + 2 \ \mathbf{\Phi^*} \overrightarrow{\nabla} \mathbf{\Phi} \right) \cdot \overrightarrow{\nabla} \Er   \quad ,       \\
     \label{em-ernst}   \left( \textsf{Re} \ \Er + | \mathbf{\Phi} |^2 \right) \nabla^2 \mathbf{\Phi}  &=& \left( \overrightarrow{\nabla} \Er + 2 \ \mathbf{\Phi^*} \overrightarrow{\nabla} \mathbf{\Phi} \right) \cdot \overrightarrow{\nabla} \mathbf{\Phi} \quad .
\eea
The complex electromagnetic and gravitational Ernst potentials, for the metric (\ref{LWP-metric}) and electromagnetic vector (\ref{Am}), are respectively defined by
\beq \label{def-Phi-Er} 
       \mathbf{\Phi} := A_t + i \tilde{A}_\varphi  \qquad , \qquad \qquad     \Er := f - \mathbf{\Phi} \mathbf{\Phi}^* + i h  \quad ,
\eeq
where $\tilde{A}_t$ and $h$ can be inferred from the following definitions\footnote{Note that the signs of the vector products depends on the order of the three-dimensional basis, and may differ from the other literature. We have chosen the following ordering ($\overrightarrow{e}_\r , \overrightarrow{e}_\varphi , \overrightarrow{e}_z$).}
\bea
    \label{A-tilde-e} \overrightarrow{\nabla} \tilde{A}_\varphi &:=&  \frac{f}{\r} \overrightarrow{e}_\varphi \times (\overrightarrow{\nabla} A_\varphi + \omega  \overrightarrow{\nabla} A_t ) \ \ , \\
    \label{h-e}    \overrightarrow{\nabla} h &:=& - \frac{f^2}{\r} \overrightarrow{e}_\varphi \times \overrightarrow{\nabla} \omega - 2 \ \textsf{Im} (\mathbf{\Phi}^*\overrightarrow{\nabla} \mathbf{\Phi} )  \ \ .
\eea
The Ernst equations (\ref{ee-ernst-ch})-(\ref{em-ernst}) are invariant under a $SO(2,1)$ group of Lie-point transformations. One of those is the Ehlers transformation
\beq \label{ehlers}
\Er \longrightarrow \bar{\Er} = \frac{\Er}{1+ic\Er} \qquad \quad ,  \quad  \qquad  \mathbf{\Phi} \longrightarrow  \bar{\mathbf{\Phi}} = \frac{\mathbf{\Phi}}{1+ic \Er} \quad .
\eeq
This map is able to move a chosen seed (i.e. a metric and electromagnetic vector potential which fulfil the equations (\ref{field-eq-g})-(\ref{field-eq-A})) described by the complex Ernst potentials $(\Er,\mathbf{\Phi})$, in the space of solutions of the theory, to get an inequivalent physical solution determined by a new doublet of Ernst potentials $(\bar{\Er},\bar{\mathbf{\Phi}})$. Therefore the Ehlers transformation is an effective way to generate new solutions starting from known seed spacetimes, without integrating the equations of motion, which are not trivial being a system of coupled partial differential equations. In particular this kind of Ehlers transformation have been recognised to be able to add the NUT parameter to a generic axisymmetric and stationary metric written in the LWP form (\ref{LWP-metric})\footnote{Note that there are other uses of the Ehlers transformations which are able to embed the seed spacetime in a swirling rotating background, see \cite{swirling},\cite{marcoa-removal}}; see for instance \cite{reina-treves}, \cite{enhanced} for adding the gravitomagnetic mass to Kerr or Kerr-Newman cases.\\

\section{Plebanski-Demianski with a double NUT parameter}
\label{sec:generation}

We want to add the NUT parameter to the accelerating, dyonically charged and rotating black hole solution,
therefore the first step we have to carry out consists in casting the full Plebanski-Demianski \cite{Plebanski-Demianski} seed into the LWP ansatz (\ref{Am})-(\ref{LWP-metric}). In terms of the spherical-like coordinates $(r,x =\cos \theta)$\footnote{So the range of angular coordinates are $x\in[-1,1]$ and $\varphi \in [0,2\pi]$.} the LWP metric takes the form
\beq \label{lwp-rx}
        ds^2 = -f(r,x) \left[ dt - \omega(r,x) d\varphi \right]^2 + \frac{1}{f(r,x)} \left[ e^{2\gamma(r,x)}  \left( \frac{{d r}^2}{\Delta_r(r)} + \frac{{d x}^2}{\Delta_x(x)} \right) + \rho^2(r,x) d\varphi^2 \right] \ \ .
\eeq
The Plebanski-Demianski metric is determined by eq. (\ref{lwp-rx}) with the following seed functions\footnote{It is worth clarify that the $\gamma$ function considered here and in (\ref{lwp-rx}) does not coincide exactly with the one in (\ref{LWP-metric}). Here, for brevity, $\gamma$ absorbs also part of the coordinate transformation $(\r,z) \to (r,x)$. Since just the metric functions $f$ and $\omega$ vary through the Ehlers transformation there is no loss of generality in doing so, and considering (\ref{lwp-rx}) as the LWP metric in $(r,x)$ coordinates is more concise and it has the advantage of avoid to deal with $z(r,x)$. Note also that $\hat{\om}$ is just a constant and have not to be confused with the functions of the LWP metric $\om(r,x)$ or $\bar{\om}(r,x)$ characterising the rotation of the black hole before and after the Ehlers transformation.}
\bea
        f(r,x)&:=& \frac{\hat{\om}^2 \D_x -\D_r}{\hat{\om} \Om^2  \mathcal{R}^2 }  \ \ , \\
        \om(r,x)&:=& \frac{\hat{\om}(r^2 \D_x + x^2 \D_r)}{\D_r-\hat{\om}^2\D_x} \ \ , \\
        \g(r,x)&:=& \mezzo \log \left( \frac{\D_r-\hat{\om}^2\D_r}{\Om^4} \right)  \ \ , \\
        \rho(r,x)&:=& \frac{\sqrt{\D_r} \sqrt{\D_x}}{\hat{\om} \Om^2} \ \ , \label{rho} \\
        \Delta_r(r)&:=& -\hat{\om} (e^2 + p^2 + k \hat{\om}^2) + 2 m \hat{\om} r - \e \hat{\om} r^2 + 2 \hat{n} \a r^3 + k \a^2 \hat{\om} r^4   \ \ ,   \\
        \Delta_x(x)&:=& -k \hat{\om} - 2 \hat{n} x + \e \hat{\om} x^2 - 2 m \a \hat{\om} x^3 + \a^2\hat{\om}(e^2 + p^2 + k \hat{\om}^2) x^4  \ \ , \\
        \Omega (r,x) &:=& 1 - \a r x \ \ , \\
        \mathcal{R}(r,x)&:=& \sqrt{r^2+\hat{\om}^2 x^2} \ \ , \label{pd-end}
\eea 
whereas the electromagnetic potential is specified by 
\bea \label{PD-A}
       A_t(r,x) &:=& - \frac{e r + \hat{\om} p x}{\mathcal{R}^2} \ \ , \\
       A_\varphi(r,x) &:=&  \frac{e \hat{\om} r x^2 -  p x r^2}{\mathcal{R}^2} \ \ .
\eea
From the definitions of the Ernst complex potentials (\ref{def-Phi-Er})-(\ref{h-e}) it's easy to deduce\footnote{It might be useful to remember the form of the gradient in the spherical-like coordinates $(r,x)$: $\overrightarrow{\nabla} f \propto \sqrt{\D_r} \overrightarrow{e}_r \p_r f + \sqrt{\D_x} \overrightarrow{e}_x \p_x f $.}
\bea
      h(r,x) &=& \frac{2 \big\{n r + \hat{\om} \big[ m x -kr^2\a-x^2 \a (e^2+p^2+k\hat{\om}) \big] \big\}  }{\Om \mathcal{R}^2} \  \  , \\
      \tilde{A}_\varphi (r,x) &=& \frac{e x \hat{\om} - p r}{\mathcal{R}^2} \  \  , \\
      \Er (r,x) &=& \frac{r \hat{\om} \D_x + i\big\{ \Om^2 \hat{\om} \big[ -ikr\hat{\om} + x (e^2+p^2+k\hat{\om}^2) \big] +x\D_y\big\}}{rx \Om^2 \hat{\om}(-ir+\hat{\om}x) }   \  \  ,  \\
      \mathbf{\Phi} (r,x) &=& - \frac{e + ip}{r+i\hat{\om}x}  \  \  .
\eea

The transformed Ernst potentials $(\bar{\Er},\bar{\mathbf{\Phi}})$ are build thanks to the Ehlers transformation (\ref{ehlers}), they read
\bea
\bar{\Er}(r,x) &=& -\frac{i \big[ \Om^2\hat{\om} (-ikr\hat{\om} +x (e^2+p^2+k\hat{\om}^2)\big] - i r \hat{\om} \D_x +x\D_y}{\Om^2 \hat{\om} \big[ -i c k r \hat{\om} + x r (ir-x\hat{\om}) + cx(e^2+p^2+k\hat{\om}^2)\big]-icr\hat{\om} \D_x+cx\D_r} \ , \\ 
\bar{\mathbf{\Phi}}(r,x) &=&  \frac{(p-ie)xr\Om^2\hat{\om}}{\Om^2 \hat{\om} \big[ -i c k r \hat{\om} + x r (ir-x\hat{\om}) + cx(e^2+p^2+k\hat{\om}^2)\big]-icr\hat{\om} \D_x+cx\D_r} \  \ .
\eea
Those complex functions basically represent the Plebanski-Demianski solution with an additional NUT charge, and by construction are solution of the Ernst  equations (\ref{ee-ernst-ch})-(\ref{em-ernst}). In case one wants to write this solution in the form of a metric and gauge potential, we have just to use the definitions (\ref{def-Phi-Er})-(\ref{h-e}), to obtain the line element\footnote{A Mathematica notebook with this metric can be found in the arXiv source folder, for the reader convenience. The PD seed (\ref{lwp-rx})-(\ref{PD-A}) also contained there, just by imposing $c=0$.}
\beq \label{lwp-rx-new}
        ds^2 = - \frac{f(r,x)}{|1+ic\Er|^2} \left[ dt - \bar{\omega}(r,x) d\varphi \right]^2 + \frac{|1+ic\Er|^2}{f(r,x)} \left[ e^{2\gamma(r,x)}  \left( \frac{{d r}^2}{\Delta_r(r)} + \frac{{d x}^2}{\Delta_x(x)} \right) + \rho^2(r,x) d\varphi^2 \right] \ \ .
\eeq
Note that $f$ rescales by a factor $|1+ic\Er|^2$, the $\g(r,x)$ function remains invariant under the Ehlers transformation, while 
\beq
     \bar{\om} = \om(r,x) + \om_0 + c^2 \ \frac{r^2\D_r(\D_x+k\Omega^2\hat{\om})^2 + x^2 \D_x [\D_r+\Om^2 \hat{\om}(q^2+k\hat{\om}^2)]^2}{x^2 r^2\hat{\om} \Om^4 (\D_r-\hat{\om^2}\D_x)} \ +
     \eeq
     \beq
               2 c \frac{r^3\a \D_r (\D_x+k\Om^2\hat{\om}) + x \Om (\D_r-\D_x\hat{\om}^2)[\D_r+\Om\hat{\om}(q^2+k\hat{\om}^2-2mr)] + x r \hat{\om}  \{ x\a\D_r [\D_x\hat{\om} +\Om^2(q^2+k\hat{\om}^2)] \}}{x r^2 \a\hat{\om} \Om^2 (\D_r-\hat{\om^2}\D_x)}\nn .
\eeq
The non-null component of the electromagnetic vector, up to an additive constant, results 
\bea \label{barAt_new}
 \bar{A}_t &=&  \frac{xr\hat{\om}\Om^2 \big\{ cer\hat{\om} \D_x + cpx\D_r + \hat{\om} \big[ -xr(er+px\hat{\om}) +c (pqx+ker\hat{\om} +kpx \hat{\om}^2) \big] \Om^2 \big\} }{|1+ic\Er|^2} \\
 \bar{A}_\varphi &=& \bar{A}_{\varphi_0} + \frac{er\hat{\om}\D_x+px\D_r}{\hat{\om}^2\D_x-\D_r} + c  \ \frac{(er+px\hat{\om})\D_x\D_r+\hat{\om}[px\hat{\om}(q^2+k\hat{\om}^2)\D_x+ker\D_r]\Om^2}{x r\hat{\om} \Om^2 (\D_r-\hat{\om^2}\D_x)} - \bar{\om} \bar{A}_t  \nn \ .
\eea
$\bar{A}_{\varphi_0}$ is an arbitrary constant that can be set to remain with a regular magnetic potential on the symmetry axis (when the magnetic charge is not present, otherwise is not possible to remove the Dirac string) and $q:=\sqrt{e^2+p^2}$. $\om_0$ is another arbitrary constant which generally defines the angular speed of the asymptotic observers and therefore can be used to move the position of a string-like angular momentum density such as the Misner string. \\

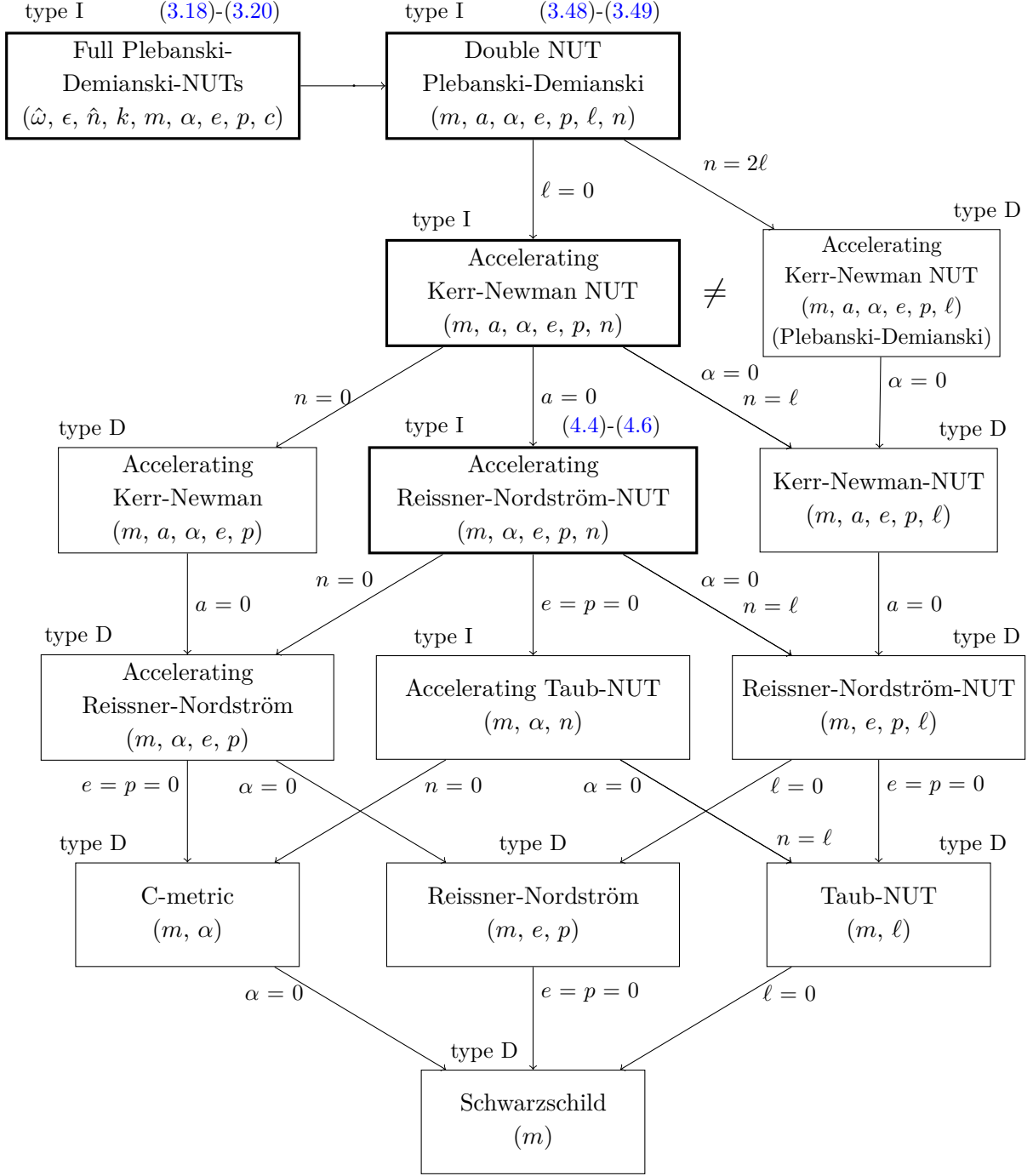
\tikzstyle{rect} = [draw,rectangle, fill=white!20, text width =3cm, text centered, minimum height = 1.5cm,scale=0.9]
\begin{figure}[H]
\vspace{-1cm}

\begin{center}

\begin{tikzpicture}
\node[rect,scale=1.2,anchor=south,label=above:{type I \quad \qquad (\ref{lwp-rx-new})-(\ref{barAt_new})},text width=4cm,line width=1.2pt](Full PD NUTs){Full Plebanski-Demianski-NUTs \hspace{0.1cm}\\ ($\hat{\om}$, $\epsilon$, $\hat{n}$, $k$, $m$, $\alpha$, $e$, $p$, $c$)};
    \node[rect,scale=1.2,anchor=south,label=above :{type I \quad \qquad (\ref{new-para-inizio})-(\ref{new-para-fine})},text width=4cm,, right of= Full PD NUTs,node distance=5.5cm,line width=1.2pt](PD double nut){Double NUT \qquad \qquad Plebanski-Demianski ($m$, $a$, $\alpha$, $e$, $p$, $\ell$, $n$)};
    \node[rect,scale=1.2,anchor=north,label=above left:{type I},label=right:{\; $\mathlarger{\mathlarger{\mathlarger{\neq}}}$},text width=4cm,below of=PD double nut,node distance=3cm,line width=1.2pt](acc kerr newman nut){Accelerating \qquad \qquad Kerr-Newman NUT ($m$, $a$, $\alpha$, $e$, $p$, $n$)};
    \node[rect,scale=1.1,anchor=north,label=above right:{type D},text width=3.5cm, right of= acc kerr newman nut,node distance=5.5cm](PD){Accelerating \qquad Kerr-Newman NUT \qquad ($m$, $a$, $\alpha$, $e$, $p$, $\ell$) (Plebanski-Demianski)};
    \node[rect,scale=1.2,anchor=north,label=above :{type I \qquad \qquad (\ref{acc-rn-nut})-(\ref{RN-Aphi})},text width=4.5cm,below of=acc kerr newman nut,node distance=3cm,line width=1.2pt](acc RN NUT){Accelerating \qquad \qquad \qquad
    Reissner-Nordström-NUT \qquad \qquad \qquad($m$, $\alpha$, $e$, $p$,  $n$)};
    \node[rect,scale=1.2,anchor=north,label=above left:{type D},text width=3.5cm, left of= acc RN NUT,node distance=5cm](acc kerr newman){Accelerating Kerr-Newman  \qquad \qquad ($m$, $a$, $\alpha$, $e$, $p$)};
    \node[rect,scale=1.2,anchor=north,label=above right:{type D},text width=3.2cm, right of= acc RN NUT,node distance=5cm](kerr newman nut){Kerr-Newman-NUT ($m$, $a$, $e$, $p$, $\ell$)};
    \node[rect,scale=1.2,anchor=north,below of=acc RN NUT,node distance=3cm,label=above left:{type I},text width=4.3cm](acc NUT){Accelerating Taub-NUT \qquad \qquad \qquad \qquad ($m$,  $\alpha$, $n$)};
    \node[rect,scale=1.2,anchor=north,right of=acc NUT,node distance=5cm,label=above right:{\; type D},text width=4cm](RN NUT){Reissner-Nordström-NUT \qquad \qquad ($m$, $e$, $p$, $\ell$)};
    \node[rect,scale=1.2,anchor=north,left of=acc NUT,node distance=5cm,label=above left:{type D \;},text width=4cm](acc RN){Accelerating \qquad \qquad Reissner-Nordström \qquad \qquad ($m$, $\alpha$, $e$, $p$)};
    \node[rect,scale=1.2,anchor=north,below of=acc RN,node distance=3cm,label=above left:{type D},text width=3cm](acc){C-metric \qquad \qquad ($m$, $\alpha$)};
    \node[rect,scale=1.2,anchor=north,below of=acc NUT,node distance=3cm,label=above:{type D},text width=4cm](RN){Reissner-Nordström \qquad \qquad ($m$, $e$, $p$)};
    \node[rect,scale=1.2,anchor=north,below of=RN NUT,node distance=3cm,label=above right:{type D},text width=3cm](NUT){Taub-NUT \qquad \qquad ($m$, $\ell$)};
    \node[rect,scale=1.2,anchor=north,below of=RN,node distance=3cm,label= above: {type D \qquad \qquad \qquad},text width=3cm](Schwarz){Schwarzschild \qquad ($m$)};

\draw[->] (Full PD NUTs) -- node [right] { . }(PD double nut);
\draw[->] (PD double nut) -- node [right] {$\ell$ = 0}(acc kerr newman nut);
\draw[->] (acc kerr newman nut) -- node [right] {$a$ = 0} (acc RN NUT);
\draw[->] (PD double nut) -- node [right,near start] {\; \; \;$n=2\ell$}(PD);
\draw[->] (acc kerr newman nut) -- node [right, near start] {\; \; $\alpha$ = 0}(kerr newman nut);
\draw[->] (acc kerr newman nut) -- node [right] {\; \; $n$ = $\ell$}(kerr newman nut);
\draw[->] (acc kerr newman nut) -- node [left] {\;$n$ = 0}(acc kerr newman);
\draw[->] (PD) -- node [right,near start] {$\alpha$ = 0} (kerr newman nut);
\draw[->] (kerr newman nut) -- node [ right] {$a$ = 0} (RN NUT);
\draw[->] (acc RN NUT) -- node [right] {$e$ = $p$ = 0}(acc NUT);
\draw[->] (acc kerr newman) -- node [right] {$a$ = 0}(acc RN);
\draw[->] (acc RN NUT) -- node [right,near start] {\; \; $\alpha$ = 0} (RN NUT);
\draw[->] (acc RN NUT) -- node [right] {\; \; $n$ = $\ell$} (RN NUT);
\draw[->] (acc RN NUT) -- node [left,near start] {$n$ = 0 \;\;} (acc RN);
\draw[->] (acc RN) -- node [left,near start] {$e$ = $p$ = 0}(acc);
\draw[->] (acc RN) -- node [left,near start] {$\alpha$ = 0 \;} (RN);
\draw[->] (acc NUT) -- node [right,near start] {\; $n$ = 0}(acc);
\draw[->] (acc NUT) -- node [left,near start] {$\alpha$ = 0 \;}(NUT);
\draw[->] (acc NUT) -- node [right,near end] {\; \;$n$ = $\ell$}(NUT);
\draw[->] (RN NUT) -- node [right,near start] {\; $\ell$ = 0} (RN);
\draw[->] (RN NUT) -- node [right,near start] {$e$ = $p$ = 0} (NUT);
\draw[->] (acc) -- node [left,near start] {$\alpha$ = 0\; } (Schwarz);
\draw[->] (RN) -- node [right,near start] {$e$ = $p$ = 0} (Schwarz);
\draw[->] (NUT) -- node [right,near start] {\;$\ell$ = 0} (Schwarz);

\end{tikzpicture}
\caption{Structure of the complete family of accelerating solutions in the Einstein-Maxwell theory which contain a single black hole. The spacetimes not yet known in the literature are emphasized in a  bold line rectangle. The new family of solutions are of Petrov Type I and may contain up to two independent NUT parameters ($\ell$ and $n$). From the double NUT accelerating Kerr-Newman metric (i.e. the double nut generalization of the Plebanski-Demianski family, all other known  solutions may be retrieved as subcases of the first one, turning off some of its parameters. Be aware of the fact that whenever the acceleration is turned off, that is the accelerating horizon and conformal boundary is pushed to spatial infinity, there is no more distinction between the two NUT parameters, $\ell$ and $n$. Turning off all parameters the usual Minkowski spacetime is retrieved too.}
\label{fig:grafico}

\end{center}
\end{figure}

This solution is the most general presented in this article and it represents an extension of the Plebanski-Demianski space-time because it is always endowed with at least a NUT parameter. It possess nine explicit physical parameters ($\hat{\omega}, e , p, k, m , \epsilon, \hat{n}, \a, c$), and few more implicit parameters such as the range of the azimuthal angle $\varphi$ for instance, that can be useful to move possible conical singularities typical of the accelerating metrics\footnote{Arbitrary additive constants to the electromagnetic potential or to the rotational function of the metric $\bar{\omega}(r,x)$ that often are considered trivial gauge constants in this context may play a role to regularise some properties of the solution as the rotation speed on the axis of symmetry.}. The standard Plebanski-Demianski solution, i.e. the seed (\ref{lwp-rx})-(\ref{PD-A}), can be recovered by switching off the real parameter $c$ introduced by the Ehlers transformation. Clearly we do not expect this metric to belong to the D type according to Petrov classification. That's because even just a subcase of the above solution, such as the accelerating Taub-NUT spacetime, belongs to the I-type, as shown by \cite{Podolsky-2020}. The easiest way to obtain the accelerating Taub-NUT metric is to apply a Ehlers transformation to the uncharged C-metric, thus vanishing both electromagnetic charges ($e,p$) and rotations parameters, such as $\hat{\om}$, in the seed. Otherwise it can be obtained as a limit from the metric of section \ref{sec:PD-BH-NUTs} or as explicitly discussed in section \ref{sec:acc-RN-Nut}. \\
In any case a direct computation of the scalar invariant quantities related to the Weyl tensor can prove the algebraic class of the full nine-parameters black hole family (\ref{lwp-rx-new})-(\ref{barAt_new}). For instance by evaluating the scalar invariant 
\beq
                     I^3 -27 J^2 \ ,
\eeq
where
\begin{equation}
I = \Psi_0\Psi_4 - 4 \Psi_1\Psi_3 + 3\Psi_2^2 \qquad , \qquad
J = \det
\begin{pmatrix}
\Psi_0 & \Psi_1 & \Psi_2 \\
\Psi_1 & \Psi_2 & \Psi_3 \\
\Psi_2 & \Psi_3 & \Psi_4
\end{pmatrix} \ ,
\end{equation}
we can infer that the generated spacetime is not of type D, but algebraically general. The definition of the Newman-Penrose scalars $\Psi_i$ can be found in appendix \ref{app-PD-bh+L}. Of course some specific sub-cases, which can be obtained by vanishing some parameters may be algebraically special as the seed, for $c=0$.\\

\subsection{Accelerating Kerr-Newman black hole with two NUTs} 
\label{sec:PD-BH-NUTs}

In this section we will restrict a bit the generality of the solution (\ref{lwp-rx-new})-(\ref{barAt_new}), in order to cast the resulting metric in a more convenient form. We would like to obtain a new parametrization more intuitively related to the physical properties carried by each integration constant. In this way all the limits of the general black hole metric to its known subcases would be direct and clear. At this purpose we perform a shift and dilatation in the time and angular coordinates and a re-parametrisation, as proposed in \cite{Podolsky-2020}:
\bea \label{new-par-t}
      t &\to & t - \frac{(\ell+a)^2}{a} \varphi  \\
      \varphi &\to & - \frac{\hat{\om}}{a} \varphi \nn  \\ 
      x &\to & \frac{a(ax+\ell)}{a^2+\ell^2} \nn \\
      e &\to & e/\sqrt{S} \nn \\
      p &\to & p/\sqrt{S}  \nn  \\  
      m &\to & \frac{m +\a \frac{\ell}{\hat{\om}}(a^2-\ell^2+e^2+p^2)}{S} \nn  
\eea
\bea  
        \hat{\om} & \to &  \frac{a^2+\ell^2}{a} \nn \\
       c &\to& c S    \nn \\
       \hat{n} &\to & \frac{\ell(a^2+\ell^2)^2 + a m (\ell^4-a^4) \a}{(a^2+\ell^2)^2-2am\ell(a^2+\ell^2)\a +a^2\ell^2(a^2+e^2+p^2-\ell^2)\a^2} \nn  \\
      \e & \to &   \frac{(a^2+\ell^2)^2+4am\ell(a^2+\ell^2)\a -a^2(a^2-\ell^2)(a^2+e^2+p^2-\ell^2)\a^2}{(a^2+\ell^2)^2-2am\ell(a^2+\ell^2)\a+a^2\ell^2(a^2+e^2+p^2-\ell^2)\a^2} \nn \\
      k & \to &   \frac{a^2(a^2-\ell^2)}{(a^2+\ell^2)^2-2am\ell(a^2+\ell^2)\a +a^2\ell^2(a^2+e^2+p^2-\ell^2)\a^2}  \nn \\
      S & \to &  1+ \frac{-2am\ell(a^2+\ell^2)\a +a^2\ell^2(a^2+e^2+p^2-\ell^2)\a^2}{(a^2+\ell^2)^2} \label{new-par-S}
\eea
Having performed the above manipulations, we arrive at the following metric\footnote{A Mathematica notebook with this metric can be found in the arXiv source files, for the reader convenience.}
\bea \label{PD-bh double nut}
     \hspace{-0.7cm}  && ds^2 =  \frac{1}{ \Om^2} \left\{ - f \left( dt -  \omega d\varphi \right)^2  + \frac{1}{f} \left[ \left(Q-a^2 P \right) \left( \frac{dr^2}{Q} +\frac{dx^2}{P} \right) + \rho^2 d\varphi^2 \right] \right\} \ , \nn \\
     \hspace{-0.7cm}  &&  A_\mu = \left\{A_t, \ 0 ,\ 0 ,  A_\varphi \right\} 
\eea 
where

\bea
        Q(r) &=&(r-r_+)(r-r_-) \left[1+\frac{a r (a-\ell)\alpha}{a^2+\ell^2} \right] \left[ 1-\frac{a r (a+\ell)\alpha}{a^2+\ell^2} \right]\ \ ,  \\
        P(x)&=& (1-x^2)\bigg\{1+ \frac{a (ax+\ell) \alpha}{(a^2+\ell^2)^2}\left[-2m(a^2+\ell^2)+a(ax+\ell)(q^2+a^2-\ell^2)\alpha \right]\bigg\}\ \ , \nn \\
        \Om(r,x) &=& 1-\frac{ a r (ax+\ell)\a}{a^2+\ell^2} \ \ , \nn \\
        r_\pm &=& m \pm \sqrt{m^2+\ell^2-a^2-q^2}\label{rp} \ \ , \nn \\
         \rho(r,x) &=& \sqrt{ P \ Q} \ \ , \nn \\
        f(r,x) &=& \frac{r^2 (ax+\ell)^2 \left(Q-a^2P \right)\Om^4}{\mathcal{R}} \ \ , \nn \\
        \mathcal{R}(r,x) &=& \left\{r^2 (ax+\ell) \Om^2- c r\left[\Om^2(a^2-\ell^2)-a^2P\right] \right\}^2 \ ,\nn \\
        &+& \left\{r (ax+\ell)^2 \Om^2+ c(ax+\ell)\left[Q-\Om^2(q^2+a^2-\ell^2)\right] \right\}^2 , \nn \\
        \omega(r,x) &=& -\frac{1}{f \mathcal{R} a }\bigg[Qr^2\bigg(\Om^2(ax+\ell)^2+c(\Om^2(a^2-\ell^2)-a^2P)\bigg)^2+ \\
        &+&a^2P(ax+\ell)^2\bigg(\Om^2 r^2+c(Q-(q^2+a^2-\ell^2)\Om^2)\bigg)^2\bigg] + \nonumber\\
        &+& \frac{2c\Om^2(ax+\ell)^2}{f \mathcal{R} a^2 \alpha}(a^2+\ell^2)\Big[\big(Q-a^2P\big)(q^2 +a^2 -2mr-\ell^2)\Om^2-Q\big(Q\Om-a^2P\big)\Big] \nonumber \\
        &+& \frac{2cr(ax+\ell)\Om^2}{f \mathcal{R} a}\bigg\{Q \Big[\Om^2 \Big((a^2-\ell^2)(axr+r^2+r\ell)+(q^2+a^2-\ell^2)(a^2x^2-\ell^2)\Big) + \nonumber \\
        &-& a^2P\Big(r^2+2(ax+\ell)^2 \Big)\Big]-a^2P(r-2\ell)(ax+\ell)(q^2+a^2-\ell^2)\Om^2\bigg\} + \frac{(a+\ell)^2}{a} + \om_0\ \ , \nn 
\eea
while the resulting potentials become
\bea
        A_t(r,x) &=& \frac{r(ax+\ell)\Om^2}{\mathcal{R}} \bigg\{c \bigg[e r \Big((a^2-\ell^2)\Om^2-a^2P \Big)+ \nn \\
        & -& \ p(ax+\ell) \Big(Q-(q^2+a^2-\ell^2)\Om^2 \Big) \bigg]- r (ax+\ell) \Big(p (a x+\ell)+e r \Big)\Om^2 \bigg\} \ , \\
        A_\varphi(r,x) &=& A_{\varphi_0}- A_t  \omega + \frac{r (ax+\ell)\Om^2}{\mathcal{R} f a } \bigg\{ c \Big(p(ax+\ell)+er\Big)a^2P Q + \nn \\ 
        &+&\Big[(ax+\ell)\Big(er^2-c p (q^2+a^2-\ell^2)\Big)a^2P+ r\Big(p(ax+\ell)^2 - \ ce(a^2-\ell^2)\Big)Q\Big]\Om^2 \bigg\} \ . \ 
        \label{PD-bh double nut fine} 
\eea
This metric represents the double NUT generalization of the type D Plebanski-Demianski family. Indeed, turning off the parameter $c$, it gives us back the Plebanski-Demianski family in the same form as written in \cite{Podolsky-2021}. \\%
After an Ehlers transformation, to have the metric in a more significant physical form, usually a constant shift of the radial coordinate and a further rescaling of the solution constants are required, in a similar way done in section \ref{sec:acc-RN-Nut}, in case of null angular momentum. Then the spacetime would depend on seven explicit parameters: $m, a, e, p, \a, \ell, n$, directly related to the mass, the angular momentum, electric and magnetic charge, the acceleration and the two independent NUT parameters. Where we call $n$ the newly introduced NUT parameter, which clearly directly depends on the Ehlers parameter $c$, when $c=0$, $n=2\ell$.\\
In the rotating but non accelerating case a good example for a new parameterization\footnote{While with two NUTs others values for c might exist, when the seed has no Misner string, this parameterization for $c$ is unique to get Kerr-Newman-NUT \cite{enhanced}. Also this choice is consistent with the zero angular momentum case of section \ref{sec:acc-RN-Nut}.} is given by
\bea \label{para-PD double nut}
      && r \to  \frac{1}{\sqrt{1+c^2}} \left[ r - \frac{2c(cm-n+\ell)}{1+c^2} \right] \ \ , \hspace{1cm}  t \to t \sqrt{1+c^2}  \ \ , \hspace{0.9cm} m \to \frac{m-c^2m+2c(n-\ell)}{(1+c^2)^{3/2}} \ , \nn \\
     && \ell  \to  \frac{2cm-n+c^2(n-\ell)+\ell}{(1+c^2)^{3/2}} \ \ , \hspace{2cm} e \to \frac{e}{\sqrt{1+c^2}} \ \ , \hspace{1.1cm} p \to \frac{p}{\sqrt{1+c^2}} \  , \hspace{1cm}  \\
    && a \to \frac{a}{\sqrt{1+c^2}} \ \ , \hspace{4.4cm} \nn 
\eea 
Thanks to this parameterization it is possible to verify that in the non accelerating case the new NUT parameter $n$ couples to the $\ell$ trivially and actually it is possible to fine tune $n=\ell$ such that both annihilates and we remain completely without the NUT contribution, i.e. Kerr-Newman spacetime. In the presence of the acceleration we will see in section \ref{sec:removal} how the  functional behavior of $c$ in (\ref{para-PD double nut}) can be modified. \\
The structure of the limits to the main notable spacetimes that can be obtained from this solution is pictured in figure \ref{fig:grafico}.\\

\subsection{Dyonic Plebanski-Demianski Black Hole}

In the above setting, when the nut parameter introduced by the Ehlers transformation, $c$ is null or equvalently $n=2\ell$, we obtain a convenient form of the classic type-D Plebanski-Demianski black hole
\bea \label{PD-bh}
     \hspace{-0.7cm}  && ds^2 =  \frac{1}{ \Om^2} \left\{ - \frac{\D_r}{\mathcal{R}^2} \left[ dt- \Big( a(1-x^2) + 2\ell(1-x) \Big)d\varphi \right]^2 + \frac{\mathcal{R}^2}{\D_r}dr^2 +\frac{\mathcal{R}^2}{\D_x}dx^2 + \frac{\D_x}{\mathcal{R}^2}\bigg[a dt - \big[r^2+(a+\ell)^2\big]d\varphi \bigg]^2  \right\} , \nn \\
     \hspace{-0.7cm}  &&  A_\mu = \left\{ -\frac{er+p(ax+\ell)}{\mathcal{R}^2}  , \ 0 ,\ 0 , \ \frac{er(1-x)(a+ax+2\ell)+p(\ell/a+x)[(a+\ell)^2+r^2]}{\mathcal{R}^2} - \frac{p\ell}{a} \right\} \ ,
\eea 
where
\bea
        \Om(r,x) &=& 1-\frac{\a a r (\ell+ a x)}{a^2+\ell^2} \ \ ,  \\
        \mathcal{R}(r,x) &=& \sqrt{r^2+(\ell+ax)^2} \ \ , \\
        \D_r(r) &=& \left(r-r_+ \right) \left(r-r_- \right) \left[1 + \frac{a-\ell}{a^2+\ell^2} \ \a a r \right] \left[1 -  \frac{a+\ell}{a^2+\ell^2} \ \a a r \right]  \ \ ,   \\
        \D_x(x) &=& \left(1-x^2\right) \left[1- \frac{ \ell+ax}{a^2+\ell^2} \ \a a r_+ \right] \left[1- \frac{\ell+ax}{a^2+\ell^2}  \ \a a r_- \right]   \ \ ,  \\
        r_\pm   &=&     m \pm \sqrt{m^2+\ell^2-a^2-e^2-p^2} \ \ . \label{rp-PD-bh}
\eea
This parametrization has the advantage with respect to previous line elements to have clear all limits to its sub-cases. Precisely at this purpose the gauge constant in $A_\varphi$ has been properly chosen and the vector potential has been properly realigned thanks to the results of \cite{enhanced}. This solution resembles the one presented in \cite{Podolsky-2020}, in fact the metric is identical, but it has a fundamental difference in the vector potential. Therefore we have a different physical interpretation. In \cite{Podolsky-2020} what is interpreted there as the magnetic charge parameter can be reabsorbed in the electric charge parameter and therefore it disappears from the solution. On the other hand in our proposal the magnetic charge parameter $p$ is an independent integrating constant directly related to the magnetic monopole charge of the solution. Hence our solution, thanks to the electromagnetic field modelled by (\ref{PD-bh}), can be considered more general and really dyonic, contrary to the one in \cite{Podolsky-2020} (indeed in this latter the magnetic Reissner-Nordstrom black hole is not even contained). The same problem is also present in the cosmological generalisation of the Plebanski-Demianski black hole \cite{Podolsky-2022}. 
For sake of completeness we present in appendix \ref{app-PD-bh+L} the generalisation of the solution (\ref{PD-bh}), representing the dyonic version of the accelerating, rotating and electromagnetically charged in the presence of the  cosmological constant. \\
In order to verify that the black hole described in (\ref{PD-bh}) is actually endowed with a monopolar magnetic field we can compute the magnetic charge enclosed in a two dimensional ball which extends at spatial infinity\footnote{In the case we do not consider the asymptotic values of the magnetic field we would remain with a coordinate dependent charge, related to the radius $r$, of the integrating spherical surface $\p \S_r$: $ P_r = - [p (r^2+a^2) + 2 e \ell r + p \ell^2]/\mathcal{R} $ .} $\p \S_\infty$, for large $r$ 
\beq 
        P = \frac{1}{8\pi} \int_{\p\S_\infty} F_{\m\n} dx^\m \w dx^\n = \lim_{r \to \infty} - \frac{1}{4\pi} \int_0^{2\pi} d\varphi \int_{-1}^{1} \p_x A_\varphi dx =  p  \ \ .
\eeq
In case one might be worried about not conventional asymptotic fall off for the full solution when performing this calculation, it is safe to switch off the more problematic parameters such as the acceleration $\a$ and the NUT parameter $\ell$ to remain with a  asymptotically locally flat solution.
The same computation, executed for the solutions of \cite{Podolsky-2020} and \cite{Podolsky-2022}, gives instead a null magnetic charge and $\sqrt{e^2+p^2}$ as electric charge, that is another indication that the magnetic parameter of these metrics is just immaterial and thus can be better eliminated from these solutions. In fact the solutions discussed in \cite{Podolsky-2020} and \cite{Podolsky-2022} have not the topological defect of the gauge vector potential typical of the monopole magnetic charged solutions, the Dirac string. \\

\subsection{Removal of the Misner String from NUTty spacetimes} 
\label{sec:removal}

The metric expressed as in eqs. (\ref{PD-bh double nut})-(\ref{PD-bh double nut fine}) is also plagued by some wire-like singularity caused by the presence of the two NUT parameters, the one already located in the seed $\ell$ and the one introduced by the Ehlers transformation $c$. In fact it is well known that NUT parameters bring a nodal singularity usually called Misner String, which has bizarre physical behaviour, indeed it was considered by Misner itself barely physical or “a counterexample to almost everything” \cite{misner-counterexample}. The Misner string can be interpreted as a semi-infinite massless source of angular momentum \cite{bonnor}, i.e. a spinning rod on one symmetry semi-axis. Because this represents a  singular behaviour Misner has proposed to cure it by a periodic identification of the time-like coordinate. While this proposal fixes the geometrical issues of the NUT parameter it produces physical problems such as closed time-like curves and therefore loss of causality. One of the main motivations of this article is to study the possibility to remove some of the problematic characteristics of the NUT parameter without recur to periodic time. On the other hand we would like to study the possible constructive interaction between the two NUT parameters. First of all we recall that from a mathematical point of view one of the undesired geometrical feature caused by the gravitomagnetic charge is the irregular behaviour of the $\omega(r,x)$ function on the axis of symmetry, that is for $x=\pm1$. More specifically, the nutty metrics usually present a wire-like discontinuity on the $z-$axis, very similar to the one related to the Dirac string in the presence of magnetic monopole potentials. Thus we analyse the value of $\omega(r,x)$ both on the north and south azimuthal semi-axis respectively 
\bea
        \lim_{x \to 1} \omega (r,x) &=&          - \frac{2c(a^2+\ell^2) + a c (a-\ell)(ac-4m-c\ell) \a}{a^2\a}  + \om_0  \ \ ,\\  
        \lim_{x \to -1} \omega (r,x) &=&  4 \ell - \frac{2c(a^2+\ell^2) + a c (a+\ell)(ac+4m+c\ell) \a}{a^2\a}  + \om_0  \ \ .        
\eea
In general in the presence of NUT parameter, these two values are not equal, therefore the $\omega(r,x)$ function of the LWP metric (\ref{LWP-metric}) is not continuous on the axis of symmetry. Actually it presents a jump along the $z$-axis proportional to 
\beq \label{Deltaw}
          \D \omega =  \lim_{x \to 1} \omega (r,x) - \lim_{x \to -1} \omega (r,x) = -4\ell + 4 c ( 2 m + c \ell  ) \ .
\eeq
We can see from the above formula, as in the standard Plebanski-Demianski family of solutions (for $c=0$), that the absence of the Misner strings requires shutting down the nut parameter $\ell$ to restore the regularity of $\omega(r,x)$ and to avoid all other NUT related issues. On the other hand the new family of solution we presented in this article, because they possess two independent NUT parameters allow us to remove the Misner string from the accelerating and rotating black holes without completely turn off the seed NUT charge $\ell$ from the spacetime, at least until the acceleration and the rotational parameters, $\a$ and $a$ remains simultaneously non null. In fact we can impose the regularity of the $\omega(r,x)$ function by requiring
\beq \label{reg-c}
             \D\omega =0 \quad \Longleftrightarrow  \quad  c = \frac{-m \pm \sqrt{m^2+\ell^2}}{\ell} \ .
\eeq
On top of (\ref{reg-c}) it could be appropriate to fix a gauge for the $\omega$ in order to make manifest the absence of discontinuity also of the full metric element $g_{t\varphi}$ on the axis of symmetry
\beq \label{w0}
         \omega_0 = \frac{2c(a^2+\ell^2) + a (a-\ell)[c^2 (a + \ell) - 2\ell] \a}{a^2\a} \ \ .
\eeq
Note that for the above values of the constants $c$ and $\omega_0$ in (\ref{reg-c})-(\ref{w0}), not only the functions $\omega(r,x)$ and the $g_{t\varphi}$ are continuous on the $z$-axis but they are also null there. Furthermore the constraint (\ref{reg-c}) makes the angular velocity $g_{t\varphi}/g_{\varphi\varphi}$ constant on the all axis of symmetry, which then vanish in the non accelerating limit, as happens for C-metrics without NUT. Therefore we manage to remove the typical  Misner string features from this NUT spacetime, in the sense of the Bonnor interpretation \cite{bonnor}, the string as a (semi infinite) spinning rod of angular momentum. 
The falloff of the $g_{t\varphi}$, for large values of the radial coordinate $r$, is of zero order, as typically happens for rotating  and accelerating black holes\footnote{It can decay faster in special sub-cases such as $\ell=a$.}. \\
To incorporate better the new NUT parameter $c$ we can follow the spirit of the non-accelerating parameterization (\ref{para-PD double nut}). We need to add the transformation for the acceleration and the Ehlers NUT parameter as follows
\beq
            \a \to  \a \sqrt{1+c^2} \qquad  , \hspace{2cm} c \to \frac{-m+\sqrt{m^2+n^2-2n\ell}}{n}     \label{c-n-alpha} \ \ .
\eeq 
These values are consistent both with the non accelerating and non rotating cases of (\ref{para-PD double nut}) and (\ref{para-rn1}) respectively. They are also coherent with the non-accelerating single NUT case of \cite{enhanced}. Then, in terms of the new parameterization, the regularising condition (\ref{reg-c}) simply become $\D\om = 4(n-\ell)=0$, hence $n=\ell$. 
Note that, by inspecting the behaviour of the function which determines the black hole event horizons, $Q(r)$, with the choice of (\ref{para-PD double nut}), (\ref{c-n-alpha})\footnote{Note that the following formulas the values for the constants ($m,a,\a,e,p,\ell,n$) are considered, with a slight abuse of notation, as reparameterized according to eqs. (\ref{para-PD double nut}), (\ref{c-n-alpha}). Therefore they differs from the values pre-rescaling, as in (\ref{reg-c}), (\ref{w0}) or section \ref{sec:PD-BH-NUTs}, even though are labeled with the same characters.}, that is  
\beq
           Q(r) = (r-\bar{r}_+)(r-\bar{r}_-)\left[ 1 - \frac{a(a-\ell)(r-\hat{r}_-)\a}{a^2+\ell^2} \right] \left[ 1 + \frac{a(a+\ell)(r-\hat{r}_-)\a}{a^2+\ell^2} \right]  \ ,
\eeq
where
$$
   \bar{r}_\pm := m \pm \sqrt{m^2+(n-\ell)^2-a^2-e^2-p^2} \qquad \qquad \text{and} \hspace{1cm} \hat{r}_\pm:=m \pm \sqrt{m^2+n^2-2n\ell} \quad ;
$$
we see that the position of the inner and outer horizons, after the regularising constraints (\ref{reg-c}) or equivalently $n=\ell$, does not depend anymore from the NUT parameters $\ell$ nor $n$:
\beq \label{reg-horizon}
             \bar{r}_\pm \ \Big|_{n=\ell} \ = \ m \pm \sqrt{m^2-a^2-e^2-p^2} \ \ .
\eeq
Other relevant metric functions, which determine the conformal infinity, after the reparameterization (\ref{para-PD double nut}), (\ref{c-n-alpha}) transform as follows
\bea
         \Om(r,x) &=& 1-\frac{\a a (ax-\ell)}{a^2+\ell^2} (r-\hat{r}_-)\\
         P(x)  &=&   (1-x^2)\left\{ 1-\frac{a\a(ax-\ell)}{a^2+\ell^2} (\hat{r}_+-\hat{r}_-) + \left[\frac{a\a(ax-\ell)}{a^2+\ell^2} \right]^2 (a^2+e^2+p^2-\ell^2) \right\}
\eea
The full solution in the new parametrisation, with $n$ instead of $c$, can be found in section \ref{app:new-param}.\\
So, for vanishing acceleration parameter $\a$, we get precisely the Kerr-Newman metric, while if we take the limit of vanishing $\ell$ and $a$ (in this order) we recover the accelerating Reissner-Nordstrom-NUT of section \ref{sec:acc-RN-Nut}; finally further switching off the charges we obtain the Taub-NUT of \cite{mann-stelea-chng} exactly in the form of \cite{Podolsky-2020}. For $n=2 \ell$ we obtain the Plebanski-Demianski black hole of section \ref{sec:PD-BH-NUTs} (up to the sign of $\ell$, which can be safely switched). In case we would like to recover also the standard Kerr-Newman electromagnetic potential an extra rotation of the Faraday tensor is needed, or better it is possible to use an enhanced version of the Ehlers transformation \cite{enhanced}, as done in the case without angular momentum respectively in appendix \ref{app:electrom-duality} or in section \ref{app:enhanced}. \\
However we will see in section \ref{sec:acc-RN-Nut}, where the accelerating Reissner-Nordstrom-NUT black hole is presented, that in absence of rotation this regularisation of the Misner string can not be generically pursuit, unless ending in some known subcases belonging to the D-type, that is basically eliminating completely the NUT parameters from the accelerating solution. We recall that also in absence of acceleration a fine tuning of the two NUT parameters (for a Ehlers transformed Kerr-Newman-NUT spacetime, for example) allows one to regularise the metric in a trivial way, that is removing completely the NUT charge. \\
On the other hand, the rotating and accelerating case described by eqs. (\ref{new-para-inizio})-(\ref{new-para-fine}), is not diffeomorphic to the Plebanski-Demianski seed, for instance because the metric, also with the constraint (\ref{reg-c})-(\ref{w0}), remains algebraically more general with respect to the seed.\\

\subsection{Convenient physical parameterization for the Plebanski-Demianski-NUTs Black Holes}
\label{app:new-param}

As discussed in section \ref{sec:removal} the parameter $n$ is better suited to describe the NUT charge, instead of the constant that labels the Ehlers transformation $c$. Indeed when the solution is written in terms of $n$ the geometrical properties of the new parameter are clearer. For instance $n$ affects directly the geometry of the event horizon. The new parametrisation makes the limits to the other accelerating NUTty solutions, such as the accelerating Reissner-Nordstrom-NUT (section \ref{sec:acc-RN-Nut}) and accelerating Taub-NUT \cite{Podolsky-2020}, metrics clearer.  \\
For this reason it's worth writing the full new solution, the Type I Plebanski-Demianski spacetime with double NUT parameters in this new parametrisation. In order to keep as compact as possible the expressions we will make use of the following constants $a, \ell, e, p, \a , n, m, \hat{r}_\pm, q$, even though $\hat{r}_\pm$ and $q :=\sqrt{e^2 + p^2}$ are not independent integration constants but they are defined as above.
Thanks to the parametrisation (\ref{para-PD double nut}), (\ref{c-n-alpha}), the full solution takes the form
\beq \label{new-para-inizio}
          ds^2 = \frac{1}{\Om^2}  \left\{- f (dt - \om d\varphi )^2 + \frac{1}{f} \left[ \Big(Q-a^2P\Big) \left( \frac{dr^2}{Q} +  \frac{dx^2}{P} \right) + \rho^2 d\varphi^2 \right]\right\} \ ,  
\eeq
where\footnote{For the reader convenience also this solution is contained and checked in the Mathematica notebook uploaded between the arXiv source files.}
\bea
      \Om &=& 1 - \frac{\a a (r- \hat{r}_-)(ax-\ell)}{a^2+\ell^2}  \ , \nn \\      
      f &=& - \frac{(\hat{r}^2_-+n^2)(r-\hat{r}_-)^2(ax-\ell)^2(a^2P-Q)\Om^4}{\mathcal{R}}    \ , \nn \\ 
      Q &=&  \Big[r^2 - 2mr + q^2 + a^2 - (n-\ell)^2\Big] \left[ 1 - \frac{\a a (r-\hat{r}_-)(a-\ell)}{a^2+\ell^2} \right] \left[ 1 + \frac{\a a (r-\hat{r}_-)(a+\ell)}{a^2+\ell^2} \right] \ , \nn \\
      P &=& (1-x^2)\left\{ 1-\frac{a\a(ax-\ell)}{a^2+\ell^2} (\hat{r}_+-\hat{r}_-) + \left[\frac{a\a(ax-\ell)}{a^2+\ell^2} \right]^2 (a^2+e^2+p^2-\ell^2)  \ \right\} ,\nn \\
    \mathcal{R} &=& (ax-\ell)^2 \left\{ Q\hat{r}_- - \Big[nr(ax-\ell) + (a^2+q^2-anx+n\ell-\ell^2) \hat{r}_-   \Big] \Om^2   \right\}^2 + \nn \\ 
    &+& (r-\hat{r}_-)^2 \left\{  a^2P\hat{r}_- + \Big[ n r (\ell-ax) + (anx -a^2+\ell^2-\ell n) \hat{r}_- \Big]  \Om^2  \right\}^2  \nn \\
   A_t  &=& \frac{\sqrt{n^2+\hat{r}_-^2} (r-\hat{r}_-)(\ell-ax)\Om^2}{\mathcal{R}} \bigg\{a^2 e (\hat{r}_- - r) P \hat{r}_-  + \big[a^3 p x \hat{r}_- - (pe^2\hat{r}_- + p^3 \hat{r}_- + en(r-\hat{r}_-)^2 )\ell \big]  + \nn \\
   &+&  p (\ell-ax) Q \hat{r}_- +(np-e\hat{r}_-)(r-\hat{r}_-)\ell^2 + p\ell^3 \hat{r}_- + a^2\big[(r-\hat{r}_-)(e\hat{r}_-+npx^2)-p\ell\hat{r}_- \big] + \nn \\ 
   &+& ax\Big[e^2p\hat{r}_- + p^3\hat{r}_- +en(r-\hat{r}_-)^2 -p\ell(2nr-2n\hat{r}_- + \ell \hat{r}_-) \Big]\Om^2 \bigg\} \nn 
   \eea
   \bea
   A_\varphi &=& A_{\varphi_0} - A_t \om +   \frac{(r-\hat{r}_-)\big[np(\ell-ax)^2+e(a^2-\ell^2)\hat{r}_-\big]Q\Om^2 }{a\sqrt{n^2+\hat{r}_-^2}(r-\hat{r}_-)(\ell-ax)(a^2P-Q)\Om^2} + \nn \\
   &+& \frac{a^2 P  \big\{ [e(\hat{r}_--r)+p(\ell-ax)]Q\hat{r}_-+ (ax-\ell) [en(r-\hat{r}_-)^2+(a^2+q^2-\ell^2)p\hat{r}_-] \Om^2 \big\}  }{a\sqrt{n^2+\hat{r}_-^2}(r-\hat{r}_-)(\ell-ax)(a^2P-Q)\Om^2} \nn \\
   \om &=& w_0 +\frac{a^2P \{Q\hat{r}_--[n \hat{r}_-^2 + nr^2+\hat{r}_- (a^2-\ell^2 + q^2 - 2nr)]\Om^2\}^2}{a(n^2+\hat{r}_-^2)(r-\hat{r}_-)^2 (a^2P-Q) \Om^4 } + \nn \\
        &+& \frac{Q\{a^2P\hat{r}_- + [n(ax-\ell)^2+(\ell^2-a^2)\hat{r}_-]\Om^2 \}^2}{a(n^2 + \hat{r}_-^2)(ax-\ell)^2(a^2P-Q)\Om^4} -   \frac{2n\hat{r}_-(a^2+\ell^2) Q (Q\Om-a^2P)}{\a a^2 (n^2+\hat{r}_-^2) (r-\hat{r}_-)^2 (a^2P-Q)\Om^2} +  \nn \\
         &-&  2n\frac{(a^2+\ell^2)\{r \hat{r}_-^2 - \hat{r}_-^3 + nr(n-2\ell)+\hat{r}_-[a^2 + q^2 - (n-\ell)^2] \}}{\a a^2 (n^2+\hat{r}_-^2) (r-\hat{r}_-)^2}     +   \nn \\
          &+&  \frac{2na P \hat{r}_- \{ -[2(ax-\ell)^2 + (r-\hat{r}_-)^2 ]Q - (2\ell +r-\hat{r}_-)(ax-\ell)(a^2+q^2-\ell^2)\Om^2 \} }{(n^2+\hat{r}_-^2) (r-\hat{r}_-)(ax-\ell) (a^2P-Q)\Om^2} + \nn \\
           &+&  2n \frac{\hat{r}_- [ (r-\hat{r}_-) (a^2-\ell^2)(r-\hat{r}_- +ax-\ell) + (a^2x^2-\ell^2)(a^2+q^2-\ell^2)]Q}{a(n^2+\hat{r}_-^2) (r-\hat{r}_-)(ax-\ell) (a^2P-Q) } \ . \label{new-para-fine}
\eea
The constant $\om_0$ can be fixed arbitrary, in order to have well defined limit to all the subcases (for vanishing $\a, a, \ell, ...$), a good choice is
\beq
        \om_0 = \frac{-2n\hat{r}_-[4a^2n^2+(n^2+\hat{r}_+\hat{r}_-)^2] + a (n^2+\hat{r}_+\hat{r}_-)[n^4+\hat{r}_-^3 \hat{r}_+ + n^2\hat{r}_-(5\hat{r}_+-3\hat{r}_-) ]\a}{4 \a a^2 n^2 (n^2 + \hat{r}_-^2)} \ . \nn 
\eeq
In fact from this spacetime a large class of accelerating black holes, in a convenient physical parametrisation, for the Einstein-Maxwell theory can be easily obtained. The Plabansky-Demiansky family for $n=2\ell$, the type I accelerating Reissner-Nordstrom-NUT of section \ref{sec:acc-RN-Nut}. \\
Of course other families of accelerating black holes, belonging to Petrov type I, are known. However these others involve departing from only a Rindler background. Instead they represent Rindler background in an external (back-reacting) electromagnetic or gravitational field, as discovered by Ernst in \cite{ernst-remove} and \cite{ernst-generalized-c}, see also \cite{marcoa-pair} and \cite{multipolar-acc} for their respective generalisations. Therefore the family (\ref{new-para-inizio})-(\ref{new-para-fine}) represents the more general single black hole spacetime in an accelerating Rindler background. The inclusion of additional external fields is not technically difficult, thanks to the solution generating techniques, but makes the metric and electromagnetic potential more involved. \\

\paragraph{Kerr-Newman-NUT black hole in Rindler-NUT background} An interesting and novel sub-case is obtained from the general solution (\ref{new-para-inizio})-(\ref{new-para-fine}) switching off the NUT parmeter of the seed $\ell$ (but not the Ehlers one, so $ c  \neq 0 $ or similarly $n \neq 2\ell $). In this case we remain with an accelerating Kerr-Newman-NUT black hole embedded in a Rindler-NUT background, but with a single NUT parameter. Therefore the NUT of the Rindler-NUT background and the NUT of the Kerr-Newman-NUT black hole can not be independent as when $\ell \neq 0$. While the NUT charge is considered a property of the spacetime symmetry axis, its total value depends on all the NUT charges carried by the fundamental building block of the axisymmetric and stationary solution, which are the solitons according to the inverse scattering technique \cite{belinski-book}. When we refer to the NUT charge of black hole we mean the NUT charge encoded in the solitons that generate that specific black hole, and similarly for other horizons.\\
The $n \neq 0 , \ell = 0 $ accelerating Kerr-Newman-NUT solution has not to be confused with the usual, Type-D, accelerating Kerr-Newman-NUT black hole (which can be obtained for $n=2\ell$ and $\ell\neq0$). Indeed the novel accelerating Kerr-Newman-NUT belongs to the type I. Another difference between the two accelerating Kerr-Newman-NUT solutions comes from the fact that when turning off the angular momentum parameter $a$, the type I solution retain the accelerating black hole structure (for instance the accelerating horizons and conformal boundaries are preserved); on the contrary the type D basically loose all its initial acceleration properties. Since we have only a single NUT parameter it is not possible to erase the Misner string, unless in the trivial case $n=0$, because in this particular case, the no Misner string constraint becomes $\D\om = 4n = 0 $. This constraint also removes the last remaining NUT parameter, so we recover the standard accelerating Kerr-Newman black hole.  \\

\paragraph{Binary system interpretation}
The possibility of removing the Misner string by means of the Ehlers transformation is not completely new. This mechanism was used to remove the NUT parameter from the rotating black hole with a conformally coupled scalar field to remain with a globally asymptotically flat metric \cite{marcoa-stationary}. Also usually it is removed from binary rotating configurations by fine tuning the two NUT parameters associated with each one of the sources \cite{alekseev-belinski-kerr}. Having in mind the double black hole configuration can, indeed, help to understand the physics underlying our model. As explained in \cite{bubble} accelerating black holes can be thought as a particular limit of a binary system, where one of the event horizons is pushed at spatial infinity while keeping the distance of the two black holes finite, as drawn in the rods diagram representation of figure \ref{fig:binary}.        
\begin{figure}[h]
\centering
\begin{tikzpicture}
 

\draw[black,thin] (-10.3,-2) -- (-4.4,-2);
\draw[black,thin] (-9,-2.8) -- (-5,-2.8);
\draw[black,->] (-10.5,-3.6) -- (-4.2,-3.6);

\draw (-10.6,-2) node{$t$};
\draw (-10.6,-2.8) node{$\phi$};
\draw (-4.5,-3.8) node{$z$};
\draw (-10.2,-3.9) node{{\small $(a)$}};
\draw (-9,-3.8) node{{\small $w_0$}};
\draw (-8,-3.8) node{{\small $w_1$}};
\draw (-7,-3.8) node{{\small $w_2$}};
\draw (-6,-3.8) node{{\small $w_3$}};
\draw (-7.7,-1) node{{ {\it Two black holes}}};


\draw[gray,dotted] (-9,-2) -- (-9,-3.6);
\draw[gray,dotted] (-8,-2) -- (-8,-3.6);
\draw[gray,dotted] (-7,-2) -- (-7,-3.6);
\draw[gray,dotted] (-6,-2) -- (-6,-3.6);

\draw[black, dotted, line width=1mm] (-10.3,-2.8) -- (-10,-2.8);
\draw[black,line width=1mm] (-9.9,-2.8) -- (-9,-2.8);
\draw[black,line width=1mm] (-9,-2) -- (-8,-2);
\draw[black,line width=1mm] (-8,-2.8) -- (-7,-2.8);
\draw[black,line width=1mm] (-7,-2) -- (-6,-2);
\draw[black,line width=1mm] (-6,-2.8) -- (-5,-2.8);
\draw[black, dotted,line width=1mm] (-5,-2.8) -- (-4.4,-2.8);

\draw[black,->] (-3.5,-2.8) -- (-1.7,-2.8) node[midway, below, sloped] {{\small $w_3 \to +\infty$}};

\draw[black,thin] (-0.6,-2) -- (4,-2);
\draw[black,thin] (-0,-2.8) -- (4.5,-2.8);
\draw[black,->] (-0.7,-3.6) -- (4.5,-3.6);

\draw (-1,-2) node{$t$};
\draw (-1,-2.8) node{$\phi$};
\draw (4.3,-3.8) node{$z$};

\draw[gray,dotted] (1,-2) -- (1,-3.6);
\draw[gray,dotted] (2,-2) -- (2,-3.6);
\draw[gray,dotted] (3,-2) -- (3,-3.6);

\draw (-0.4,-3.9) node{{\small $(b)$}};
\draw (1,-3.8) node{{\small $w_0$}};
\draw (2,-3.8) node{{\small $w_1$}};
\draw (3,-3.8) node{{\small $w_2$}};

\draw[black, dotted, line width=1mm] (-0.6,-2.8) -- (-0.2,-2.8);
\draw[black,line width=1mm] (-0.1,-2.8) -- (1,-2.8);
\draw[black,line width=1mm] (1,-2) -- (2,-2);
\draw[black,line width=1mm] (2,-2.8) -- (3,-2.8);
\draw[blue,line width=1mm] (3,-2) -- (4,-2);
\draw[blue, dotted,line width=1mm] (4.1,-2) -- (4.5,-2);


\draw (1.7,-1) node{{ {\it One accelerating black hole}}};

\end{tikzpicture}
\caption{{\small $(a)$ Rod diagram for a collinear bynary system (the Bach-Weyl solution). The limit $w_3 \to +\infty$ gives the single accelerating black hole $(b)$. Black segments on the time-like line represent event horizons while the blue line the accelerating horizon.}}
\label{fig:binary}
\end{figure}
While the picture refers to diagonal Weyl solutions, usually it is used as a basis to add rotation with the inverse scattering technique \cite{belinski-book}. That method generically can add, for the solitons of each black hole of the static binary system, a couple of parameters related to the angular momentum and the NUT charge. According to this representation we understand\footnote{Explicit computation for the diagonal case has been done in \cite{bubble}, while for the stationary case the work is yet in progress, therefore this interpretation is what we expect to find, thanks to the insight of the inverse scattering methods.} that the seed we considered to generate our new solution, the standard Plebanski-Demianski metric, can be thought as the limit of the binary system, where only the solitons associated to one of the black holes (the left one in fig. \ref{fig:binary} (a)) were endowed with rotation and NUT parameters while the solitons associated to the other, whose horizon is for $\r=0$ and $z \in [w_2,w_3]$, had both turned off or, at most, they are not independent from the black hole ones\footnote{This does not mean that the angular velocity of the seed accelerating horizon is null, but that the angular velocity depends on the physical parameters of the black hole. If we remove the black hole in the seed metric (with its associated charges), the angular velocity of the Rindler horizon vanishes, because there is no frame dragging generated by the collapsed star. An analogous behaviour can be found in the next section: when we have two independent NUT parameters, one carried by the black hole and the other carried by the Rindler horizon, then if we turn off the one of the (seed) black hole ($\ell$), still we can remain with a rotating (type I) Rindler-NUT background, as in section \ref{sec:rindler-NUT}. But if we do not have the NUT parameter $n$, we are in the Petrov D class. When the intrinsic charges of the (seed) black hole are switched off ($m=\ell=a=e=p=0$), then we remain just with a diagonal Rindler metric.}. The action of the Ehlers transformation consists in adding to the whole system a further NUT charge. Thus the solutions we have generated in section \ref{sec:generation} can be thought as the limit where the right black hole horizon is pushed to infinity as in picture \ref{fig:binary} $(b)$. Note that after the Ehlers transformation the nut charge of the left black hole is the sum of the initial quantity plus the one carried by the Ehlers transformation and similarly for the right black hole, which before the Ehlers transformation has not its own NUT parameter, but after the Ehlers transformation has acquired some non-null amount of NUT charge. Therefore, after taking the limit, we are left with the left black hole with a certain amount of NUT charge, which is not necessarily the unique contribution to the whole spacetime.  The regularisation of $\omega(r,x)$ practically consists in tuning the NUT charge brought in by the Ehlers transformation ($n$ or $c$) in order to cancel the one present into the seed ($\ell$). Hence we can remove the Misner string from the black hole, but inevitably we cannot remove all the NUT influence from the whole space-time because there is still NUT reminiscent of the Ehlers transformed right black hole, also after taking the limit described above. It is now more explicit why it is necessary the rotation in order to regularise $\omega(r,x)$: without the rotation the accelerating seed cannot have the intrinsic nut parameter ($\ell$) turned on, therefore it cannot be balanced with the extra nut. The parameterization (\ref{para-PD double nut}), (\ref{c-n-alpha}) which gives rise to the metric (\ref{new-para-inizio})-(\ref{new-para-fine}) can further clarify the physical picture. For instance it makes apparent that the {\it NUTs} contributions (where with {\it NUTs} we mean the double NUT parameter) even disappear from the event horizon, remaining with a $\bar{r}_\pm$ as the Kerr-Newman, as shown in eq (\ref{reg-horizon}). \\
This interpretation opens to the possibility of further generalisations of the present model including additional features to the Rindler background such as electromagnetic charges and angular momentum. Recently in \cite{charged-I} some generalisation of this setting were achieved with the help of the Harrison transformation. There it is also demonstrated that type I accelerating black hole stems from a limit of charged binary system, this supports the interpretation here proposed.  \\

\subsection{No black hole: Rindler-NUT background}
\label{sec:rindler-NUT}

When the physical parameters of the black hole ($m, e, p, a, \ell$) of section \ref{sec:generation} or \ref{sec:PD-BH-NUTs} are switched off, before the reparameterisation (\ref{para-PD double nut}) and (\ref{c-n-alpha}), we remain with a two parameter solution: the acceleration $\a$, and the NUT parameter introduced by the Ehlers transformation $c$, as follows
\bea \label{alpha-c-metric}
      ds^2  &=& - \hat{f}(r,x) \left[dt - \left(\frac{2\a cr^2(1-x^2)}{(1-\a rx)^2} + \om_0 \right) d\varphi \right]^2  \\ 
            &+&  \frac{1}{\hat{f}(r,x)} \left\{ \frac{1}{(1 - \a rx)^4} \left[ dr^2+ (r^2-\a^2r^4) \left( \frac{dx^2}{1-x^2} + (1-x^2) d\varphi^2 \right) \right] \right\} \ ,\nn
\eea
with 
\beq
       \hat{f}(r,x) := \frac{(1-\a^2r^2)(1-\a rx)^2}{c^2(1-\a^2r^2)^2 + (1-\a rx)^4}  \ . \nn
\eeq
This metric has not NUT charge, even though it can be generated by the Ehlers transformation, that's because the Ehlers map does not generate the NUT charge or the Misner string, just rotate the mass into the gravitomagnetic mass and vice-versa, as will be shown at the end of section \ref{app:enhanced}. Thus setting to zero all the black hole constants erase also the Misner string, as can be understood from eq. (\ref{Deltaw}). For $c=0$ the metric (\ref{alpha-c-metric}) reduces to the Rindler spacetime, and for $\a=0$ to Minkowski. This metric still belong to Petrov type D class, therefore it may stems from a limit of the Plebanski-Demianski metric. \\
On the other hand if we turn off the black hole parameters after the reparameterisation (\ref{para-PD double nut}) and (\ref{c-n-alpha}) we get another spacetime, characterised by the parameters $\a$ and $n$, which can be better suited to be considered as the background of accelerating and NUTty black holes because its limits for vanishing $n$ and $\a$ are respectively Rindler and Taub-NUT. It can be written easily from the metric of section \ref{sec:acc-RN-Nut} imposing $m=e=p=0$, it reads
\bea
                ds^2   &=& \frac{1}{-[1-\a x (r+n)]^2} \Bigg\{ \check{f}(r,x) (dt - \check{\om}(r,x) d\varphi)^2 + \frac{1}{\check{f}(r,x)} \bigg[ dr^2 +  \\
                      &+& [1-\a^2(r+n)^2] \left(\frac{(r^2-n^2) \ dx^2}{(1-x^2)(1-2nx\a)} + (r^2-n^2) (1-x^2)(1-2nx\a) d\varphi^2 \right) \bigg] \Bigg\} \ \ ,  \nn 
\eea
with 
\bea
            \check{f}(r,x)    &=&  \frac{2(r^2-n^2)[1-(r+n)^2 \a^2][1-x\a(r+n)]^4}{(r+n)^2 (1-x\a(r+n))^4+(r-n)^2[1-\a^2(r+n)^2]} \ , \nn \\ 
             \check{\om}(r,x)  &=&  - \frac{\a x^2(r-3n)(r+n)-(r+n)^2\a+2nx(1+(r+n)^2\a^2)}{[1-\a x(r+n)]^2} \ .
\eea
Indeed this metric represents the superposition of the massless Taub-NUT and the Rindler spacetimes. In this case the Petrov class remains general, as the black hole, i.e. type I. That's because after the reparameterisation the ``old'' mass parameter is not zero, since it is partially contained in the ``new'' parameter of the Ehlers transformation $n$, through eq. (\ref{c-n-alpha}). This is one of the main differences between the actual parameter of the Ehlers transformation $c$ and its new form $n$. \\

\section{No angular momentum ($ \ell = 0, \ a = 0 $): 
    Accelerating dyonic Reissner-Nordstrom-NUT black hole}
\label{sec:acc-RN-Nut}

When the angular momentum parameter $a$ in (\ref{PD-bh double nut})-(\ref{PD-bh double nut fine}) is set to zero also the acceleration vanishes, therefore this is not a good limit for describing an accelerating and electromagnetically charged black hole with NUT. This is an issue related to the parametrization choice of the Plebanski-Demianski black hole that can be encountered also in case one would like to use, as a seed for the Ehlers transformation, the (\ref{PD-bh})-(\ref{rp-PD-bh}) metric directly with $a=0$\footnote{Alternatively to avoid this problem it is possible to use as a seed the general Plebanski-Demianski metric (\ref{lwp-rx})-(\ref{pd-end}) and switch off $\hat{\omega}$ properly to remain as a seed with a charged C-metric.}. On the other hand if, in the accelerating Kerr-Newman solution with double NUT parameters (\ref{PD-bh double nut})-(\ref{PD-bh double nut fine}), we first set the intrisic nut parameter $\ell$ to zero and, subsequently, we take the limit of vanishing rotational parameter $a$, we remain with an accelerating Reissner-Nordstrom black hole with the NUT parameter provided by the Ehlers transformation (encoded in $c$ or $n$).\\
In order to have a clear form for the metric we have to shift the radial coordinate and rescale the time and some constant as follows
\bea \label{para-rn1}
       r &\to& \sqrt{\frac{\tilde{r}_+}{\tilde{r}_+-\tilde{r}_-}} (r - \tilde{r}_-)  \quad , \qquad e \ \to \ \sqrt{\frac{\tilde{r}_+}{\tilde{r}_+-\tilde{r}_-}} \ e  \quad , \qquad  p \ \to \ \sqrt{\frac{\tilde{r}_+}{\tilde{r}_+-\tilde{r}_-}} \ p \ \ ,  \\
     t &\to& \sqrt{\frac{\tilde{r}_+-\tilde{r}_-}{\tilde{r}_+}} \ t  \hspace{1.4cm} , \qquad c \ \to \ \frac{\sqrt{-\tilde{r}_+ \tilde{r}_-}}{\tilde{r}_+} \quad \hspace{0.4cm} , \qquad  \a \ \to \  \sqrt{\frac{\tilde{r}_+-\tilde{r}_-}{\tilde{r}_+}}  \ \a \ \ , \\
    m &\to& \mezzo \sqrt{\tilde{r}_+^2 - \tilde{r}_+\tilde{r}_-} \hspace{1.1cm} , \qquad \tilde{r}_\pm := m \pm \sqrt{m^2+n^2} \ \ . \label{para-rn2}
\eea 
Then the metric takes the form\footnote{A Mathematica notebook with this metric can be found in the arXiv source files, for the reader convenience.}

\bea \label{acc-rn-nut}
     \hspace{-0.7cm}  && ds^2 =  \frac{1}{ \Om^2} \left[ - \frac{\D_r \Om^4}{\mathcal{R}} \big( dt + \omega d\varphi \big)^2 + \frac{\mathcal{R}}{\D_r \Om^4}dr^2 +\frac{\mathcal{R}}{\D_x \Om^4}dx^2 + \frac{\mathcal{R}\D_x}{\Om^4} d\varphi^2  \right] ,
\eea 
where
\bea \label{RN-Omega}
        \Om(r,x) &=& 1 - \a x (r - \tilde{r}_-)  \ \ ,  \nn \\
         \D_r(r) &=& \big[ (r-\tilde{r}_+)(r-\tilde{r}_-) + e^2 + p^2 \big] \big[ 1 - \a^2  (r-\tilde{r}_-)^2 \big] \ \ , \nn \\
        \D_x(x) &=& (1-x^2) \big[ 1 - \a x (\tilde{r}_+ - \tilde{r}_-) + \a^2 x^2 (e^2 + p^2) \big]  \ \ ,  \nn \\
        \mathcal{R}(r,x)  &=&  \frac{\tilde{r}_-\big\{ r - \tilde{r}_+ +2 \a x (e^2+p^2) + \big[ (e^2+p^2)x^2 + \frac{\D_y}{1-\a^2(r-\tilde{r}_-)^2}  \big] \a^2 (\tilde{r}_--r) \big\}^2 - \tilde{r}_+(r-\tilde{r}_+)^2\Om^4}{\tilde{r}_--\tilde{r}_+} \   , \nn \\
        \omega(r,x)   &=&  \frac{2\sqrt{-\tilde{r}_+\tilde{r}_-} \big\{ \D_x - (r- \tilde{r}_-) \a [ 2x - \a (r-\tilde{r}_-)] \D_x - (1-x^2) \Om^2 \big\} }{\big[ (1-x^2) \a (\tilde{r}_- - \tilde{r}_+)   \big]\Om^2} + \om_0   \ . \label{w_acc-rn-nut}
\eea
The metric is supported by a electromagnetic field whose non-null component of the vector potential are 
\bea
        A_t &=&  \frac{e\sqrt{\tilde{r}_+}(\tilde{r}_--r)\Om^2 + p\sqrt{-\tilde{r}_-} \big\{ r - \tilde{r}_+ +2 \a x (e^2+p^2) + \big[ (e^2+p^2)x^2 + \frac{\D_y}{1-\a^2(r-\tilde{r}_-)^2}  \big] \a^2 (\tilde{r}_--r) \big\}}{\sqrt{\tilde{r}_+-\tilde{r}_-} \mathcal{R}}   \ , \nn \\
        A_\varphi   &=&  \frac{e}{x} \sqrt{\frac{\tilde{r}_-}{\tilde{r}_- - \tilde{r}_+}} \left( 1-\frac{\D_x}{\Om^2} \right) + px \sqrt{\frac{\tilde{r}_+}{\tilde{r}_+ - \tilde{r}_-}} + A_t \om - A_{t0} \ \ . \label{RN-Aphi}
\eea
This solution remains of type I according to the Petrov classification, therefore it can not belong to the Plebanski-Demianski class which is algebraically special. It was first built in \cite{tesi-giova} from the accelerating RN seed. It represents the dyonically charged generalisation of the Chng-Mann-Stelea work \cite{mann-stelea-chng}, which describes an accelerating Taub-NUT metric, directly in the convenient parametrisation proposed by Podolsky and Vratny in \cite{Podolsky-2020}. The limits to the uncharged metric can be obtained easily switching off the electromagnetic charges ($e \to 0 , \ p \to 0$) and for $\om_0=0$. Also the other limits to type D known solution, such as Reissner-Nordstrom-NUT or the charged C-metric are straightforward, just vanishing the acceleration $\a$ or the NUT parameter $n$ respectively.\\
By construction this space-time presents a Misner string singularity brought in by the Ehlers-NUT parameter $n$. The $\om_0$ constant fixes the angular velocity of an asymptotic observer and can be adjusted to move the Misner string, which in general can not be removed, as done in section \ref{sec:removal} (unless in the trivial case $c=0$ or imposing periodic time coordinate). That's because the seed space-time one uses in order to obtain the accelerating RN-NUT solution (\ref{RN-Omega})-(\ref{RN-Aphi}) is not endowed with an intrinsic NUT parameter (as the $\ell$ in the rotating case). So there is not a counterpart to balance the nut carried by the Ehles transformation. In fact if we scrutinize the regularity of the metric, as done in section \ref{sec:removal}, in this case we find a discontinuity on the azimuthal axis for the $\omega(r,x)$ function, which is quantified by  
\beq
             \D\omega = \lim_{x \to +1} \omega (r,x) - \lim_{x \to -1} \omega (r,x) = 4 \sqrt{n^2}  \ . 
\eeq
Clearly the jump of the metric function $\omega(r,x)$ can be null only for the Ehlers-NUT parameter $n=0$\footnote{This value makes the Ehlers map in an identity. It is in line with the general result of section \ref{sec:generation} where $n=2\ell$, because in the general case $\ell \neq0$, but here $\ell=0$.}, so $\tilde{r}_+=\tilde{r}_-$ or equivalently $c=0$. \\
Note that the electromagnetic field through a Ehlers transformation usually is rotated, as explained in \cite{enhanced}. In case of asymptotic flatness an enhanced Ehlers transformation \cite{enhanced} can be used instead of the normal one in order to maintain the electric and magnetic field aligned as the seed. We mean that, starting with a pure electric seed, we can remain with an electric field only. In the presence of acceleration the situation is more involved, however the enhanced Ehlers transformation can be useful to remain with a predictable non accelerating limit for the electromagnetic potential also, without the need of extra rotations of the Faraday tensor.   The enhanced Ehlers transformation, besides a better treatment of the electromagnetic field, show also some advantages in the re-parameterization, so it can be of utility to work out it explicitly, as can be found in the next section \ref{app:enhanced}.\\

\subsection{Accelerating Reissner-Nordstrom-NUT black hole via the enhanced Ehlers transformation.}
\label{app:enhanced}

It could be of some utility to explicitly add the NUT parameter to the accelerating Reissner-Nordstrom black hole via the enhanced Ehlers transformation, because it works better in presence of electromagnetic fields, in particular for asymptotically flat solutions or asymptotically Taub-NUT, as shown in \cite{enhanced}. The enhanced Ehlers transformation is basically an Ehlers mapping modified in order not to mix the electric and magnetic field, while adding the NUT parameter to a given seed solution. Essentially, it is built to preserve a relaxed flat fall-off of both the gravitational and electromagnetic field, that is an asymptotic behaviour which at most mimic the electromagnetic Taub-NUT solutions. In the case under consideration the presence of acceleration could complicate the asymptotic analysis, but at least the enhanced Ehlers transformation ensures we obtain a proper limit in the vanishing acceleration case. In particular, we would like to get the Reissner-Nordstrom-NUT black hole solution from the spacetime generated in section \ref{sec:acc-RN-Nut}. Indeed, using the standard Ehlers transformation to build the accelerating RN-NUT black hole we can get, for vanishing $\a$, only the Reissner-Nordstrom-NUT metric but not its usual associated electromagnetic field.\\

The procedure to use the new transformation is identical with respect to the one carried out in \ref{sec:Ernst-Ehlers} and \ref{sec:generation} sections. The only difference is that we have to use, instead of the standard Ehlers map (\ref{ehlers}), the enhanced one, as follows 
\beq \label{enhanced-ehlers}
\Er \longrightarrow \bar{\Er} = \frac{\Er+ic}{1+ic\Er} \qquad \quad ,  \quad  \qquad  \mathbf{\Phi} \longrightarrow  \bar{\mathbf{\Phi}} = \frac{\mathbf{\Phi}(1+ic)}{1+ic \Er} \quad .
\eeq
For simplicity we will apply it to the charged c metric, which enclose the accelerating RN black hole (the full Plebanski-Demianski or Kerr-Newman case works in a similar way). In term of the LWP metric (\ref{lwp-rx}) and potential (\ref{Am}) the diagonal seed is determined by 
\bea
          \Om(r,x) &=&  1- \a x r \ ,                                           \hspace{3.1cm}   \D_r(r) = e^2+p^2 +r (-2m+2mr^2-kr^3\a^2+r\e) \ , \nn   \\
          f(r,x) &=&  \frac{\D_r(r)}{r^2 \ \Omega^2(r,x)}   \ ,           \hspace{2.6cm}  \D_x(x) = k - 2mx\a(1-x^2) - x^2[(e^2+p^2)\a^2x^2+\e] \ ,\nn \\
          \r(r,x) &=&  \frac{\sqrt{\D_r(r)} \sqrt{\D_x(x)}}{\Om^2(r,x)}\ ,  \hspace{1.4cm} A_t(r,x) \ = \ -\frac{e}{r} \ ,\nn   \\
           \g(r,x) &=& \frac{1}{2} \log \left[ \frac{\D_r(r)}{\Omega^4(r,x)} \right] \ ,         \hspace{1.4cm} A_\varphi(r,x) \ = \ -px  \ .\nn  
\eea
Then according to the definitions (\ref{def-Phi-Er}) - (\ref{h-e}) we can find $h$ and $\tilde{A}_\varphi$ to determinate a couple of complex Ernst potential for the seed 
\bea
             \Er &=&  \frac{-kr^3\a^2+(e^2+p^2)x\a(2-xr\a) - 2m(1-r^2\a^2) +r\e}{r \ \Omega^2(r,x)} \ ,  \nn  \\
             \mathbf{\Phi} &=&  \frac{-e+ip}{r} \ .
\eea
Having all the ingredients we can generate a new solution, in terms of the complex Ernst potential, with the NUT parameter by applying the enhanced Ehlers transformation (\ref{enhanced-ehlers}):
\bea
        \bar{\Er} &=& \frac{i\{2q^2x\a-2m(1-r^2\a^2) + r [\e - (q^2x^2 + k r^2)\a^2 +ic\Omega^2(r,x)]\}}{iy(1-r^2\a^2)+c[kr^3\a^2-q^2x\a(2-r^2\a^2)+m(2-2r^2\a^2)-r\e]} \ ,\nn \\
        \bar{\mathbf{\Phi}} &=&  \frac{(c-i)(e-ip)\Omega^2(r,x)}{iy(1-r^2\a^2)+c[kr^3\a^2-q^2x\a(2-r^2\a^2)+m(2-2r^2\a^2)-r\e]} \nn \ .
\eea
Following the parameterization (\ref{new-par-t})-(\ref{new-par-S}) in this case without angular momentum we impose
\beq
    \e \to 1 - e^2 + p^2  \ \ , \hspace{2.7cm} \varphi \to - \varphi \ \ , \hspace{2.7cm} k \to 1  \ \ . \label{costr-RN}
\eeq 
While the parameterization of (\ref{para-rn1})-(\ref{para-rn2}) thanks to the enhanced trasformation is simplified into
\beq
         r \to r - \tilde{r}_- \ \ , \hspace{2.2cm} m \to \frac{1}{2} (\tilde{r}_+-\tilde{r}_-) \ \ , \hspace{2.2cm} c \to \sqrt{-\frac{\tilde{r}_-}{\tilde{r}_+}} \ \ . \nn
\eeq
As in section \ref{sec:acc-RN-Nut} we define $\tilde{r}_\pm := m \pm \sqrt{m^2+n^2}$, and we remember that, as above $q:=\sqrt{e^2+p^2}$, while the asymptotic angular velocity of the rotating frame is fixed by setting $$\om_0=\frac{2\sqrt{-\tilde{r}_-\tilde{r}_+}[1-\a(\tilde{r}_+ - \tilde{r}_-)]}{\a(\tilde{r}_+ - \tilde{r}_-)} \ \ .   $$
This value guarantees  a well defined non accelerating limit and consistent result with the known sub-cases.
Finally, thanks to the definitions (\ref{def-Phi-Er}) - (\ref{h-e}) we can come back to the metric and potential notation, which respectively reads\footnote{Also this solution can be found in the Mathematica notebook between the arXiv sources files.}
\beq \label{ehanced-acc-RN-ds}
        ds^2 = -\bar{f}(r,x) \left[ dt - \bar{\omega}(r,x) d\varphi \right]^2 + \frac{1}{\bar{f}(r,x)} \left[ e^{2\gamma(r,x)}  \left( \frac{{d r}^2}{\Delta_r(r)} + \frac{{d x}^2}{\Delta_x(x)} \right) + \rho^2(r,x) d\varphi^2 \right] \ \ ,
\eeq
where
\bea
      \bar{f}(r,x) &=& \frac{(\tilde{r}_+-\tilde{r}_-) \D_r \Om^2}{\tilde{r}_+(r-\tilde{r}_-)^2\Om^4 - \mathcal{R}^2 \tilde{r}_-} \ \ , \nn \\
      \bar{\om}(r,x) &=& - 2 \sqrt{-\tilde{r}_- \tilde{r}_+}  \left\{ 1- \frac{1}{\a(\tilde{r}_+ - \tilde{r}_-)} \left[ 1 + q^2\a^2 - \D_x \frac{1-(r-\tilde{r}_-)\a[2x-(r-\tilde{r}_-)\a]}{(1-x^2) \Om^2} \right] \right\} \ \ , \nn \\
      \mathcal{R}(r,x) &=& r-\tilde{r}_+ + 2 q^2 x\a -\left[ q^2 x^2 \frac{\D_r}{1-(r-\tilde{r}_-)^2\a^2}\right] \a^2 (r-\tilde{r}_-)  \ \ , \nn \\
      \Om(r,x) &=& 1-\a x (r-\tilde{r}_-) \ , \nn \\
      \Delta_r(r) &=& [(r-\tilde{r}_+)(r-\tilde{r}_-) + q^2][1-\a^2(r-\tilde{r}_-)^2] \ , \nn \\
      \Delta_x(x) &=&  (1-x^2) [1 - \a x (\tilde{r}_+ - \tilde{r}_- - \a x q^2)] \,
\eea
and 
\bea
      \bar{A}_t(r,x) &=& \frac{\Om^2 e [\tilde{r}_+ (r-\tilde{r}_-) \Om^2 - \tilde{r}_- \mathcal{R}]}{\tilde{r}_- \mathcal{R}^2 - \tilde{r}_+ (r-\tilde{r}_-)^2 \Om^4} + \ \ \nn  \\
      &+& \frac{p \Om^2\sqrt{-\tilde{r}_+\tilde{r}_-} \left\{(\tilde{r}_+-\tilde{r}_-)[1-\a^2(r-\tilde{r}_-)^2] - \a [q^2+(r-\tilde{r}_-)^2][2x+(1+x^2)(r-\tilde{r}_-)] \right\} }{\tilde{r}_- \mathcal{R}^2 - \tilde{r}_+ (r-\tilde{r}_-)^2 \Om^4}   \ \ , \nn \\
      \bar{A}_\varphi (r,x) &=& \frac{(p\tilde{r}_--e\sqrt{-\tilde{r}_-\tilde{r}_+})\D_x+[e\sqrt{-\tilde{r}_+\tilde{r}_-} (1-x^2) + p(\tilde{r}_+ x^2-\tilde{r}_-))]\Om^2}{(\tilde{r}_--\tilde{r}_+) x \Om^2} - \bar{A}_t \bar{\om}(r,x) + A_{t0} \ \ . \ \    \label{ehanced-acc-RN-Af}
\eea

It is easy to check that the above solution has the expected limit, for the NUT parameter $n\to0$\footnote{Note that this means $\tilde{r}_-=0, \tilde{r}_+=2m$ and $c\to0$.}, to the accelerating charged Reissner-Nordstrom black hole
\bea
         ds^2 &=&  \frac{1}{(1-\a r x)^2} \left\{ - \left( 1- \frac{2m}{r} + \frac{q^2}{r^2} \right) \big(1-\a^2r^2 \big) dt^2 + \frac{dr^2}{\left( 1- \frac{2m}{r} + \frac{q^2}{r^2} \right) \big(1-\a^2r^2 \big)} + \right. \\
         & & \hspace{2cm} + \left. \frac{r^2 dx^2}{\big(1-x^2\big) \big[ 1+\a x (q^2x\a-2m)  \big]} + r^2 \big(1-x^2\big) \big[ 1+\a x (q^2x\a-2m)  \big] d\varphi^2 \right\} ,   \nn  \\
         A_\m &=& \{ -\frac{e}{r},\ 0 ,\ 0 ,\ - p x \}  \ \ ,  \nn
\eea
which is a slightly restriction of the seed, since we introduced constraints (\ref{costr-RN}) on the integrating constants in order to retrieve a clearer black hole interpretation.\\
More interestingly, the limit for null acceleration $\a \to 0$ of the solution (\ref{ehanced-acc-RN-ds})-(\ref{ehanced-acc-RN-Af}) give precisely the Reissner-Nordstrom-NUT black hole:
\bea
    \hspace{-0.8cm}  && ds^2 =  - \frac{r^2 -2mr + q^2 - n^2}{r^2+n^2} \Big[dt + 2n(1-x)d\varphi \Big]^2 + \frac{(r^2+n^2)\ dr^2}{r^2 -2mr + q^2 - n^2} + \frac{r^2+n^2}{1-x^2} dx^2 + (1-x^2)(r^2+n^2) d\varphi^2,   \nn  \\
     \hspace{-10cm} && \  A_\m = \left[ \frac{-er+pn}{r^2+n^2} , 0 , 0 , \frac{2ner(1-x)+2pn^2+px(r^2-n^2)}{r^2+n^2} \right]  \ \ .  \nn
\eea
Thanks to these limits the interpretation that the enhanced generated solution (\ref{ehanced-acc-RN-ds})-(\ref{ehanced-acc-RN-Af}) describes an accelerating Reissner-Nordstrom-NUT black hole become natural.\\
Thus we have shown, similarly to the models discussed in \cite{enhanced}, a superior behaviour of the enhanced Ehlers transformation (\ref{enhanced-ehlers}) with respect to the standard one (\ref{ehlers}), because only the first map adds the NUT parameter while preserving the character of the seed electromagnetic field and of the physical interpretation of the electromagnetic parameters: $e$ and $p$ remain the electric and magnetic monopolar charges respectively, as we have been shown in the non accelerating limit. On the contrary the standard Ehlers transformation mix up the electric and the magnetic interpretation of $e$ and $p$, which possibly have to be disentangled with new transformations or reparameterizations, see appendix \ref{app:electrom-duality}.  \\
Analogously, one might act with the enhanced transformation also for the most general seed of section \ref{sec:Ernst-Ehlers} to have a better defined limits to electromagnetic potential of the Carter-Plebanski spacetime.\\ 
The same result of remaining with a properly alignment electromagnetic field can also be obtained combining the Ehlers transformation with some other symmetries of the Ernst equations and eventually some diffeomorphisms. However, the procedure is more convoluted with respect to the use of just the enhanced map (\ref{enhanced-ehlers}), as done in this section. \\
\paragraph{Some properties of the enhanced Ehlers transformation} The enhanced Ehlers transformation can be obtained by a combination of the standard Ehlers transformation (\ref{ehlers}) and some other Lie-point symmetries of the Ernst equations, for a complete list see \cite{enhanced}. In fact it is well known that the Ernst equations (\ref{ee-ernst-ch}) - (\ref{em-ernst}) remains invariant also for the following maps
\bea
    \label{(I)}      (I):\    && \Er \longrightarrow \bar{\Er} = \l \l^* \Er  \ \quad \qquad \ ,  \qquad \qquad \mathbf{\Phi} \longrightarrow  \bar{\mathbf{\Phi}} = \l \mathbf{\Phi} \ , \\
      (II):  \ && \Er \longrightarrow \bar{\Er} = \Er + i \ b \qquad  \ \ \ , \qquad  \qquad \mathbf{\Phi} \longrightarrow  \bar{\mathbf{\Phi}} = \mathbf{\Phi} \quad . \nn 
\eea
It is easy to check that the enhanced Ehlers transformation (\ref{enhanced-ehlers}) can be derived as the following composition  
\beq
           (II) \circ (I) \circ (\ref{ehlers}) \  \Bigg|_{\l= 1 + i c  \atop \hspace{-0.35cm} b = c}  = \ \   (\ref{enhanced-ehlers}) \ \ .
\eeq
Thus, even if more laborious, the electromagnetic field after an Ehlers transformation could be possibly adjusted also with a proper composition of $(II)$ and $(I)$. Actually, since the $(II)$ consists in just a diffeomorphism symmetry for the Einsten-Maxwell solution, it can  be sufficient to preserve the asymptotic behaviour of the fields, by using only $(I)$ (after the standard Ehlers map), in a proper gauge
, as done in \cite{enhanced}. We leave some details of this procedure in appendix \ref{app:electrom-duality}. Similarly this can be extended to the solution with angular momentum of section \ref{sec:generation} to check that the electromagnetic field coincides with the Kerr-Newman-NUT one in the vanishing acceleration limit.\\
To better understand the action of the enhanced Ehlers symmetry we can show how it transforms the Ernst potential of a relaxed asymptotically flat spacetime. With relaxed we mean that while the Ernst potentials decay for large values of the radial coordinate, it still can be possible to have Taub-NUT behaviour. This kind of gravitational and electromagnetic potentials can be written asymptotically as
\bea
\label{complex-asym} 
       \Er &\sim & 1- \frac{2 \left(M - i B\right)}{r}  + O \left(\frac{1}{r^2}\right) \quad ,  \\
    \label{complex-asym2}   \mathbf{\Phi} &\sim &  \frac{Q_e + i Q_m}{r} + O\left(\frac{1}{r^2}\right)     \quad ,
\eea
where $M, B, Q_e, Q_m$ correspond to the physical characteristic quantities of the black hole: mass, NUT, electric and magnetic charge. The enhanced Ehlers (\ref{enhanced-ehlers}) transforms the Ernst potentials (\ref{complex-asym})-(\ref{complex-asym2}) into
\bea \label{M'B'}
  \bar{\Er} &\sim &  1- \frac{2(\bar{M}-i\bar{B})}{r}  + O \left(\frac{1}{r^2}\right) = 1 - \frac{2(1-c^2)M-4Bc}{(1+c^2)r}  + i \ \frac{2(1-c^2)B + 4Mc}{(1+c^2)r} + O \left(\frac{1}{r^2}\right) , \quad  \\
  \bar{\mathbf{\Phi}} &\sim &  \frac{Q_e + i Q_m}{r} + O\left(\frac{1}{r^2}\right) \ .
\eea
In practice the enhanced Ehlers transformation turn the mass $M$ into the NUT parameter $B$ without affecting the electric $Q_e$ and magnetic $Q_m$ charges. When $c=1$ the swing is complete and the gravitomagnetic mass is completely shifted into the gravitational source and viceversa the NUT parameter B becomes the actual mass of the black hole. Thus note that, in the case under consideration, it is necessary to start with a massive solution to add the NUT parameter to seed without Misner strings. \\
To better understand why we interpret the enhanced Ehlers map as a rotation is sufficient to consider the transformation the mass and the gravitomagnetic mass $M, B$ undergo passing from (\ref{complex-asym}) to (\ref{M'B'}), which can be summarised by the operator $Rot$
\beq
          Rot = \begin{pmatrix}  \displaystyle 
 \frac{1-c^2}{1+c^2} & \displaystyle  -\frac{2c}{1+c^2} \vspace{2mm} \\
         \displaystyle  \frac{2c}{1+c^2} & \displaystyle  \frac{1-c^2}{1+c^2} 
\end{pmatrix} \ .
\eeq
When $Rot$ acts on the mass vector $v=(M,B)$, where its first component is the mass and the second component the gravitomagnetic mass, we can exactly reproduce the map between  (\ref{complex-asym}) to (\ref{M'B'}):
\beq
                Rot (M,B) = (\bar{M},\bar{B})=\left(\frac{(1-c^2)M-2Bc}{1+c^2} , \frac{(1-c^2)B + 2Mc}{1+c^2} \right) \ .
\eeq
Defining $c := \tan \psi/2$ we make apparent that the matrix $Rot$ is actually a standard rotation in the plane 

\beq
          Rot \ \bigg|_{c \to \tan \psi/2} \ = \ \begin{pmatrix}  
 \cos \psi &  -\sin \psi \vspace{1mm} \\
           \sin \psi &  \cos \psi 
\end{pmatrix} \ . 
\eeq
\\
Therefore the enhanced Ehlers transformations acts similarly to the transformation $(I)$ of eq (\ref{(I)}), but instead of rotating the electric into the magnetic charge\footnote{A direct example of how the $(I)$ symmetry rotates the electric in magnetic charge can be found in appendix \ref{app:electrom-duality}.}, it rotates the mass into its complex dual, in the Ernst formalism: the gravitomagnetic mass. The Hodge operator, acting on the Faraday tensor, produces the electromagnetic duality in four-dimension,  switching the electric in magnetic field and viceversa. The action of the Hodge operator, in this context is a discrete transformation, which can be obtained for a particular value of che cmplex constant $\l$ in the $(I)$ transformation. This observation allows us to define a gravitational Hodge operator, at least for axisymmetric and stationary electrovacuum solutions, as the enhanced Ehlers transformation with $c=1$, which is a discrete symmetry of the Ernst equations. \\
Finally it's interesting to note that the enhanced Ehlers transformation has the notable property of containing another special discrete symmetry of the Ernst qquations,  the inversion transformation (also known as Buchdahl, for null electromagnetic fields)
\beq  \label{inv}
      (inv): \ \qquad \quad   \Er  \longrightarrow \bar{\Er} = \frac{1}{\Er}   \quad , \qquad  \mathbf{\Phi} \longrightarrow  \bar{\mathbf{\Phi}} = \frac{\mathbf{\Phi}}{\Er}  \quad ,
\eeq
just as a limit, for large values of the continuous parameter $c$, that is
\beq
       \lim_{c\to \infty}  (\ref{enhanced-ehlers})       =  (inv) \ \ . 
\eeq
\\

\section{Summary and Comments}

In this article we have shown how to add, through the Ehlers transformation, the NUT parameter to the whole family of the (eight parameter) Plebanski-Demianski solutions. The novel family of electrovacuum metrics, consisting with nine non-trivial physical parameters, are algebraically more general (Type-I) with respect to the standard PD class, which belong to the type D of the Petrov classification. Remarkably this new family describes a very wide class of black holes including novel ones, such as the accelerating Kerr-Newman-NUT and the accelerating Reissner-Nordstrom-NUT spacetimes. As a known sub-case we can obtain the accelerating Taub-NUT metric obtained by Chng-Mann-Stelea \cite{mann-stelea-chng}. \\
Interestingly we note that  the ``NUT charge'' introduced by the Ehlers transformation directly affects the full spacetime and it is not only limited to the black hole as in the non accelerating context. \\
The NUT parameter usually brings into space-time axial singularities more harming than the conical ones, when the time coordinate is thought not periodic. On the other hand when the time coordinate is considered periodic there are no conical singularities, but closed timelike curves (therefore time-machines) appear near the axis, compromising the causality and therefore the physical plausibility of the model. For these reasons one of the main motivations to study NUTty solutions remains, in our opinion, the discovery of a mechanism to remove the Misner string without completely eradicating other physical properties of the solution, which may come with the NUT parameters.  \\
For this purpose we considered the full Plebanski-Demianski metric as seed, because when the metric is both accelerating and rotating it can also bring a NUT parameter. Then we apply an Ehlers transformation to the seed in order to add an extra NUT parameter. In non-accelerating settings, the NUT parameter brought in by the Ehlers transformation can not be considered a really independent integration constant, since it can be reabsorbed in the intrinsic NUT charge of the seed (whether it is present). Conversely, in the general accelerating and rotating case we have presented here, the new parameter introduced by the Lie-point transformation is completely independent and it gives rise to a different solution with respect to the seed.  The differences of the generated solution with respect to the seed are not only in the Petrov classification, but can also be intuitively understood from the fact that the additional NUT parameter survives also in the (accelerating) non-rotating limit, contrary to what happens to the NUT parameter of the original PD seed. The independence of the two NUT parameters can be useful to erase certain pathologies typical of the presence of the Misner string, at least from certain sectors of the metric, in a similar way as the Ehlers transformation revealed to be useful to remove the conical singularity from the rotating C-metric through the gravitational spin-spin coupling \cite{marcoa-removal}. Indeed, we discovered that the discontinuity on the function $\om(r,x)$ and the metric component $g_{t\varphi}$ on the axis of symmetry can be regularised. This might open a more phenomenological scenario for NUTty black holes. In fact, features that cause these kinds of divergences in the fields are usually less plausible to be observed in nature. That's the case for magnetic monopoles that produce Dirac string-like singularities, geometrically related with the Misner strings brought in by the NUT parameter.  \\
Our results not only explain how to systematically and efficiently obtain NUTty solution such as the accelerating Taub-NUT black hole found by Chng, Mann and Stelea \cite{mann-stelea-chng}, but our outcomes also clarify the nature of the NUT charge involved in these accelerating space-times, which can be obtained by the Ehlers transformation. The rotational contribution due to the NUT parameter in fact is not only intrinsic to the main black hole, but the Ehlers transformation adds NUT charge also to the Rindler background, which can be thought, in this picture, as large event horizon limit of a companion black hole of a binary system, as in picture \ref{fig:binary}.\\ 
Furthermore, the superior behaviour of an enhanced variation of the Ehlers transformation has been discussed. We have shown explicitly as the a new version of the Ehlers map, as proposed in \cite{enhanced}, preserves by construction the boundary conditions of an asymptotically flat or Taub-NUT seed, both for the metric and (especially) for the electromagnetic potential, therefore is more efficient. In fact, using the enhanced map we can retrieve the limits to all the known sub-cases without further transformations. We have also shown how the enhanced Ehlers transformation can be built by a chain composition of others known Lie-point transformations and that can be interpreted as a rotation of the standard mass into the gravitomagnetic mass, or NUT charge. Also the improved Ehlers contains interesting direct limits to some discrete transformations, such as the Buchdahl map. \\
Moreover, since a particular limit of our solution does not coincide with the recent works of \cite{Podolsky-2021} - \cite{Podolsky-2022}, where an improved form of the Plebanski-Demianski black hole is pursuit, we clarify an issue about the electromagnetic field therein indicated. We propose an alternative electromagnetic potential which is able to model and take into account consistently also the magnetic field interpretation, i.e. we write the full dyonic case of the new parameterization of the PD black hole. \\
The procedure here discussed can be generalised in various directions, first of all can be interesting to add extra external self-interacting fields as the Melvin electromagnetic universe \cite{ernst-remove} or the swirling universe of \cite{swirling}-\cite{marcoa-removal} to try to remove also the residual conical singularity usually carried by the NUTty metrics. Also the generalisation of these metrics to the presence of scalar fields can be pursuit straightforwardly, because the integrability and the solution generating techniques based on the symmetry of the Ernst equations have been successfully extended in \cite{marcoa-embedding} and \cite{marcoa-stationary}. The scalar fields we are referring to are not only the minimally coupled ones, but also all the ones related with the minimally coupling by a conformal transformation, such as conformally coupled scalar fields\footnote{Works in this direction are already in progress \cite{adolfo}.}, some types of Brans-Dicke or $f(R)$ theories, etc. Finally, note that the procedure here presented does not applies necessarily to black holes, it can be applied to wormholes or more in general to any axisymmetric and stationary seed in four-dimensions.  \\

\section*{Acknowledgements}
{\small This article is based on the results of the bachelor thesis~\cite{tesi-giova} by G.B.
\ We thank Adriano Viganò for interesting discussions on the subject, Jiří Podolský and anonymous JHEP referee for useful comments. M.A. would like to thank the Universidad Austral de Chile (UACh) and the Centro de Estudios Cientificos (CECs) for the hospitality while part of this work was done. A Mathematica notebook containing the main solutions presented in this article can be found in the arXiv source folder. 
}\\
\\

\appendix

\section{The complete Plebanski-Demianski black hole with cosmological constant}
\label{app-PD-bh+L}

In the presence of the cosmological constant the Ehlers transformation is not known at the moment, and actually finding it seems to be not an easy task due to the loss of symmetry of the Ernst equations as detailed in \cite{asto-lambda}, hence accelerating black holes with nut parameter cannot be build as done in section \ref{sec:generation}.\\
Nevertheless, for sake of completeness, we provide the reader with the most general black hole solution belonging to the type D of the Petrov classification. Of course the solution is a subclass of the Plebanski-Demianski spacetime, but here the parameterization is very well suited to describe a black hole configuration, that is all the parameters appearing in the  have a proper defined physical meaning  and therefore all the limits to the notable black hole subcases are clearly achievable just by switching off the undesired physical characteristic of the black hole. The metric was already discussed in detail in \cite{Podolsky-2021} and \cite{Podolsky-2022} with or without the cosmological constant, but its interpretation in both cases seems incorrect to us. In fact, while it is claimed in these references that the solution contains all black holes of type D and to possess magnetic charge, we cannot find traces of dyonic solutions, for instance it is not contained there neither the Reissner-Nordstrom solution with magnetic charge. In fact what is addressed as magnetic charge in  \cite{Podolsky-2021} and \cite{Podolsky-2022} is just a redundant parameter that can be gauged away from the whole solution, incorporated by a rescaling in the electric charge (or even directly set to zero). This is basically because the electromagnetic potential considered in \cite{Podolsky-2021} and \cite{Podolsky-2022} is not sufficiently general to support the presence of a consistent magnetic parameter.\\
Instead we would like to provide an extended vector electromagnetic potential and electric field which allows us to interpret the complete solution as truly magnetic. Only this latter can be considered the most general type D black hole solution in general relativity (eventually coupled with Maxwell electromagnetic field and cosmological constant). The form of the metric\footnote{A Mathematica notebook with this solution can be found in the arXiv source files, for the reader convenience.} and the gauge potential are the same of the null cosmological constant case, given in (\ref{PD-bh}), we rewrite here for completeness 
\bea
     \hspace{-0.7cm}  && ds^2 =  \frac{1}{ \Om^2} \left\{ - \frac{\D_r}{\mathcal{R}^2} \left[ dt- \Big( a(1-x^2) + 2\ell(1-x) \Big)d\varphi \right]^2 + \frac{\mathcal{R}^2}{\D_r}dr^2 +\frac{\mathcal{R}^2}{\D_x}dx^2 + \frac{\D_x}{\mathcal{R}^2}\bigg[a dt - \big[r^2+(a+\ell)^2\big]d\varphi \bigg]^2  \right\} , \nn \\
     \hspace{-0.7cm}  &&  A_\mu = \left\{ -\frac{er+p(ax+\ell)}{\mathcal{R}^2}  , \ 0 ,\ 0 , \ \frac{er(1-x)(a+ax+2\ell)+p(\ell/a+x)[(a+\ell)^2+r^2]}{\mathcal{R}^2} - \frac{p\ell}{a} \right\} , \nn
\eea 
where
\bea
        \Om(r,x) &=& 1-\frac{\a a r (\ell+ a x)}{a^2+\ell^2} \ \ , \nn \\
        \mathcal{R}(r,x) &=& \sqrt{r^2+(\ell+ax)^2} \ \ , \nn \\ 
        r_\pm   &=&     m \pm \sqrt{m^2+\ell^2-a^2-e^2-p^2} \ \ . \nn
\eea
The contribution of  the cosmological constant comes only in the functions 
\bea
       \D_r(r) \hspace{-0.2cm} &=& \hspace{-0.2cm} (r-r_+)(r-r_-)\left( 1 + \frac{a-\ell}{ a^2 + \ell^2 } \a a r \right) \left( 1 - \frac{a+\ell}{a^2+\ell^2} \a a r \right) - \frac{\L r^2}{3} \left( r^2+2\a a \ell r \frac{a^2 - \ell^2}{a^2+\ell^2} + a^2 + 3\ell^2 \right) , \nn \\
       \D_x(x) \hspace{-0.2cm} &=& \hspace{-0.2cm} (1-x^2) \left\{ 1 -2 \left( \frac{m a \a}{a^2 + \ell^2} - \frac{\L}{3} \ell \right)(\ell+ax) + \left[ \frac{a^2 \a^2 r_+ r_-}{(a^2+\ell^2)^2} +\frac{\L}{3} \right] (\ell + a x)^2 \right\} \ . 
\eea
The seven explicit independent physical parameter of the black hole are $m,a,\ell,e,p,\a,\L$, that is the parameters related to the mass, the angular momentum per unit mass, the gravitomagnetic mass, the electric charge, the magnetic charge, the acceleration and the cosmological constant respectively. A further non-trivial parameter may be considered hidden in the range of the azimuthal coordinate $\varphi$ and it can be related to possible conical defect or excess. Note that $r_\pm$ does not identify the event horizon when the cosmological constant is $\L\ne0$. \\
It is worth for this solution to report the Newman-Penrose scalars because in \cite{Podolsky-2021}, \cite{Podolsky-2022} are not correct. The problem  is not only in the missing magnetic part in the Faraday tensor, but also their result does not fulfil one of the Einstein-Maxwell equations written in terms of the Newman-Penrose formalism, i.e. 
\beq \label{eq-scalar-enp}
          \Phi_{11} = 2 \ \Phi_1 \bar{\Phi}_1  \ .      
\eeq
The following null tetrad is used 
\bea \label{tetrad}
         \bf{k} &=& \frac{\Omega(r,x)}{\sqrt{2} \mathcal{R}(r,x)} \left\{ \frac{1}{\sqrt{\D_r(r)}} \Big[ (r^2+(a+\ell)^2) \ \p_t + a \ \p_\varphi \Big] + \sqrt{\D_r(r)} \ \p_r \right\} \\
         \bf{l}  &=& \frac{\Omega(r,x)}{\sqrt{2} \mathcal{R}(r,x)} \left\{ \frac{1}{\sqrt{\D_r(r)}} \Big[ (r^2+(a+\ell)^2) \ \p_t + a \ \p_\varphi \Big] - \sqrt{\D_r(r)} \ \p_r \right\}  \\
         \bf{m}   &=& \frac{\Omega(r,x)}{\sqrt{2} \mathcal{R}(r,x)} \left\{ \frac{1}{\sqrt{\D_x(x)}} \Big[ \big(a(1-x^2) + 2\ell(1-x)\big) \ \p_t +  \p_\varphi \Big] + i \sqrt{\D_x(x)} \ \p_x \right\}
\eea
where we are considering the quantities defined in (\ref{PD-bh})-(\ref{rp-PD-bh}). Clearly all the scalar products between these vectors are vanishing apart  $ k_\m l^\m =-1$ and $\ m_\m \bar{m}^\m =1 $.\\
The algebraic structure of the Weyl tensor is determined by its ten real independent components \cite{Griffiths:2009dfa}
\bea
        \Psi_0 &:=& C_{\m\n\s\r} k^\m m^\n  k^\s m^\r \ , \nn  \\
        \Psi_1 &:=& C_{\m\n\s\r} k^\m l^\n  k^\s m^\r \ ,  \nn  \\
        \Psi_2 &:=& C_{\m\n\s\r} k^\m m^\n  \bar{m}^\s l^\r \ ,  \\
        \Psi_3 &:=& C_{\m\n\s\r} l^\m k^\n  l^\s \bar{m}^\r \ , \nn \\
        \Psi_4 &:=& C_{\m\n\s\r} l^\m \bar{m}^\n  l^\s \bar{m}^\r  \ . \nn
\eea
Among those the only non vanishing scalar is 
\beq
        \Psi_2 = \frac{\Om^3}{[r-i(ax+\ell)]^3} \left\{ (i \ell-m) \left( 1+i\a a \frac{a^2-\ell^2}{a^2+\ell^2} \right) + i \frac{\L}{3} \ell (a^2-\ell^2) + (e^2+p^2)\frac{1+\frac{\a a}{a^2+\ell^2}\big[ax(r-i\ell)-i\ell^2\big]}{r+i(ax+\ell)} \right\} . \nn
\eeq
The ten independent components of the Ricci tensor are encoded, in terms of the Newman-Penrose tetrad basis (\ref{tetrad}), into four real scalars ($\Phi_{00},\Phi_{11},\Phi_{22},R$) and three complex scalars ($\Phi_{10},\Phi_{20},\Phi_{21}$). They are all null apart the Ricci scalar $R=4\L$ and 
\beq \label{Phi11}
                      \Phi_{11} : = \frac{1}{4} R_{\m\n} (k^\m l^\n + m^\m \bar{m}^\n) = \frac{e^2+p^2}{2} \frac{\Om^4}{\mathcal{R}^4}   \ .
\eeq

The Faraday 2-form tensor which contains the electromagnetic contribution to the energy momentum tensor can be derived from (\ref{PD-bh}), through  $F := d A$, and it has the following non null components 

\bea
         F_{tr}  &=&  \frac{2r(pax+er+p\ell)-e[r^2+(ax+\ell)^2]}{[r^2+(ax+\ell)^2]^2}\\
         F_{tx}  &=&  \frac{a[2(ax+\ell)(pax+er+p\ell)-p(r^2+(ax+\ell)^2)]}{[r^2+(ax+\ell)^2]^2}\\
         F_{\varphi r} &=& \frac{(1-x)(a+ax+2\ell)[a^2ex^2-2pr\ell+2ax(e\ell-pr)+e(\ell^2-r^2)]}{[r^2+(ax+\ell)^2]^2}  \\
         F_{x \varphi} &=& \frac{[r^2+(a+\ell)^2][a^2 p x^2+2er\ell+2ax(er+p\ell)+p(\ell^2-r^2)]}{[r^2+(ax+\ell)^2]^2} 
\eea
The six independent elements of the Maxwell field strength are described by three complex Newman-Penrose scalars
\beq
       \Phi_0 := F_{\m\n} k^\m m^\n  \ \ , \qquad  \Phi_1 := \mezzo F_{\m\n} (k^\m l^\n + \bar{m}^\m m^\n)  \ \ , \qquad \Phi_2 := F_{\m\n} \bar{m}^\m l^\n \ ,
\eeq 
of which the only non-null is 
\beq
      \Phi_1   = - \frac{(e-ip)\Om^2(r,x)}{2[r + i (ax-\ell)]^2} \ .   
\eeq
This value exactly combines with (\ref{Phi11}) to fulfill the equation of motion (\ref{eq-scalar-enp}).\\
Since the only non-vanishing Newman-Penrose Weyl scalar is $\Psi_2$, ${\bf k}$ and ${\bf l}$ are double-degenerate principal null directions. This assures that when the nut parameter introduced by the Ehlers transformation is null the metric remains of Petrov algebraic type D.\\
The geometric structure of this spacetime is studied extensively in \cite{Podolsky-2021} and \cite{Podolsky-2022}, most details remain valid also for this dyonic generalisation. \\

\section{How to realign the electromagnetic field through the eletro-magnetic rotation (I)}
\label{app:electrom-duality}

Here we clarify the steps necessary in order to adjust the electromagnetic field after a standard Ehlers transformation (\ref{ehlers}). But first of all we stress that if instead of (\ref{ehlers}) one uses the enhanced Ehlers map (\ref{enhanced-ehlers}) this procedure is already built in the transformation, thus superfluous, as discussed in section \ref{app:enhanced}. We take the accelerating Reissner-Nordstrom-NUT as an example, but the machinery can be similarly applied also in the more general case with angular momentum treated in sec \ref{sec:generation} to obtain the expected limit to the Kerr-Newman-NUT solution in the vanishing acceleration case.\\
The dyonically charged C-metric-NUT was obtained in section \ref{sec:acc-RN-Nut} thanks to the standard Ehlers transformation  (\ref{ehlers}). This Ehlers transformation has the defect of mixing the electric and magnetic character of the vector gauge potential, and thus of the Maxwell field, as explained in \cite{enhanced}. In case one wants to reestablish the normal behaviour of the electromagnetic field a further transformation can be applied after the Ehlers transformation: the electromagnetic rotation duality $(I)$, as in eq. (\ref{(I)}). Since we do not want to change the overall net electromagnetic charge, we chose a particular unitary class of the transformation $(I)$, that is $\l=\exp (i d)$, where $d$ is a real constant that label the one parameter of the transformation.\\
In the dyonic case we are considering, this duality rotation turns out to have the remarkably simple effect of rotating, in the vector potential of section \ref{sec:acc-RN-Nut}, the monopolar electric and the magnetic charge parameters as follows
\begin{equation}
\left(
\begin{matrix}
e  \\
p 
\end{matrix}
\right)
\longrightarrow
\begin{pmatrix}
\cos d & \sin d \\
-\sin d & \cos d 
\end{pmatrix}
\left(
\begin{matrix}
e  \\
p 
\end{matrix}
\right)  \ ,
\end{equation}
where 
\beq \label{d}
      \arccos{d} = \frac{e^2-p^2+\frac{2 e p \tilde{r}_-}{\sqrt{-\tilde{r}_+ \tilde{r}_-}}}{(e^2+p^2)\sqrt{\frac{\tilde{r}_+-\tilde{r}_-}{\tilde{r}_+}}} \ \ .
\eeq
This choice of $d$ is the one that allows us to interpret $e$ and $p$ as the electric and magnetic charges of the spacetime, at least in the absence of acceleration. In fact in the null acceleration limit the electromagnetic vector potential, up to addictive gauge constants becomes
\beq
      A_\m = \left[  \frac{-er+p n}{r^2 + n^2} , 0 , 0 , \frac{-pxr^2+2e(1-x)rn-p(2-x)n^2}{r^2+n^2} \right] \ \ ,
\eeq
which precisely coincides with the Reissner-Nordstrom-NUT electromagnetic potential.
Note that before taking the limit, it  could be convenient to set a specific value for the free integration constant in (\ref{w_acc-rn-nut}), that is
\beq
          \om_0 = - 2 \sqrt{-\tilde{r}_+ \tilde{r}_-} \ \ ,
\eeq
in order to simplify the identification with the Reissner-Nordstrom-NUT solution, because in this way both spacetimes are described in the same rotating frame. The unitary duality transformation we used above leaves completely invariant the metric, as should be because it already possessed the correct limit to RN-NUT for vanishing acceleration. \\
The specific angle value $d$ that keep the electromagnetic field aligned with the one of the seed, when $\a$ is null, depends on the specific solution under consideration.\\  
Finally note that the unitary map $(I)$ is not really adding a new integration constant to the seed solution; it is just rearranging the electromagnetic charges $e$ and $p$ in the vector gauge potential. That's only because the solution is dyonic, so the integration constants for this kind of monopolar field are already saturated. In case the solution is purely electric or purely magnetic, then the unitary transformation $(I)$ has the effect of rotating the electromagnetic parameters and adding a new physical feature to the solution. For instance $(I)$ can transform a purely electric solution in a dyonic one\footnote{Unless, of course, the $d$ is not a special case consisting of semi-integer multiples of $\pi/2$, because in that case the map acts as a Hodge operator on the Faraday tensor, switching a electric field in a magnetic one and viceversa. In that trivial case, by applying $(I)$, we remain with the same number of integration constants for the electromagnetic vector potential.}.  \\


\newpage










\begin{thebibliography}{99}



\bibitem{ernst2}
  F.~J.~Ernst,
  {\it ``New Formulation of the Axially Symmetric Gravitational Field Problem. II''},
  Phys.\ Rev.\  {\bf 168} (1968) 1415. \href{https://doi.org/10.1103/PhysRev.168.1415}{\tt [doi:10.1103/PhysRev.168.1415]}

\bibitem{enhanced}
  M.~Astorino,
  {\it ``Enhanced Ehlers Transformation and the Majumdar-Papapetrou-NUT Spacetime''},
  \href{https://doi.org/10.1007/JHEP01(2020)123}{JHEP \textbf{01} (2020), 123} ; 
   \href{https://arxiv.org/pdf/1906.08228}{\tt [arXiv:1906.08228 [gr-qc]]}.

\bibitem{ernst-remove}
  F.~J.~Ernst,
  {\it ``Removal of the nodal singularity of the C-metric''}, 
   \href{http://scitation.aip.org/content/aip/journal/jmp/17/4/10.1063/1.522935}{J. Math. Phys. {\bf 17}, 515 (1976)}.

\bibitem{ernst-generalized-c} 
  F.~J.~Ernst,
  {\it ``Generalized C-metric''},
   \href{https://doi.org/10.1063/1.523896}{J.\ Math.\ Phys.\  {\bf 19}, 1986-1987 (1978).}

\bibitem{marcoa-pair}
  M.~Astorino,
  {\it ``Pair Creation of Rotating Black Holes''},
  \href{https://doi.org/10.1103/PhysRevD.89.044022}{Phys. Rev. D \textbf{89} (2014) no.4, 044022};
  \href{https://arxiv.org/pdf/1312.1723.pdf}{\tt [arXiv:1312.1723~[gr-qc]]}.

\bibitem{multipolar-acc}
  M.~Astorino and A.~Vigan\`o,
  {\it``Many accelerating distorted black holes''},
  \href{https://doi.org/10.1140/epjc/s10052-021-09693-6}{Eur. Phys. J. C \textbf{81} (2021) no.10, 891};
  \href{https://arxiv.org/pdf/2106.02058.pdf}{\tt [2106.02058 [gr-qc]]}.

\bibitem{misner-counterexample}
  C. W. Misner,  
  {\it ``Taub-NUT space as a counterexample to almost anything''}, 
  Relativity theory and astrophysics 1 (1967): 160.\\  \href{https://ntrs.nasa.gov/api/citations/19660007407/downloads/19660007407.pdf}{\tt [https://ntrs.nasa.gov/api/citations/19660007407/downloads/19660007407.pdf]}

\bibitem{Plebanski-Demianski}
  J.~F.~Plebanski and M.~Demianski,
  {\it ``Rotating, charged, and uniformly accelerating mass in general relativity''},
  \href{https://doi.org/10.1016/0003-4916(76)90240-2}{Annals Phys. \textbf{98} (1976), 98-127}.

\bibitem{Podolsky-2020}
J.~Podolsky and A.~Vratny,
{\it ``Accelerating NUT black holes''}, 
 \href{https://doi.org/10.1103/PhysRevD.102.084024}{Phys. Rev. D \textbf{102} (2020) no.8, 084024}; 
\href{https://arxiv.org/pdf/2007.09169.pdf}{\tt [arXiv:2007.09169 [gr-qc]]}. 

\bibitem{mann-stelea-chng}
B.~Chng, R.~B.~Mann and C.~Stelea,
{\it ``Accelerating Taub-NUT and Eguchi-Hanson solitons in four dimensions''},
 \href{https://doi.org/10.1103/PhysRevD.74.084031}{Phys. Rev. D \textbf{74} (2006), 084031} ;
\href{https://arxiv.org/pdf/gr-qc/0608092.pdf}{\tt [arXiv:gr-qc/0608092]}

\bibitem{reina-treves}
  A.~Reina and A. Treves
  {\it ``NUT-like generalization of axisymmetric gravitational fields''} ,  \href{https://doi.org/10.1063/1.522614}{Journal of Mathematical Physics 16, 834 (1975)}.

\bibitem{swirling}
M.~Astorino, R.~Martelli and A.~Vigan\`o,
{\it ``Black holes in a swirling universe''},
\href{https://doi.org/10.1103/PhysRevD.106.064014}{Phys. Rev. D \textbf{106} (2022) no.6, 064014} ;
\href{https://arxiv.org/pdf/2205.13548}{\tt [arXiv:2205.13548 [gr-qc]]}.

\bibitem{marcoa-removal}
M.~Astorino,
{\it ``Removal of conical singularities from rotating C-metrics and dual CFT entropy''}
\href{https://doi.org/10.1007/JHEP10(2022)074}{JHEP \textbf{10} (2022), 074};
\href{https://arxiv.org/pdf/2207.14305}{\tt [arXiv:2207.14305 [gr-qc]]}.


\bibitem{Podolsky-2021}
J.~Podolsky and A.~Vratny,
{\it ``New improved form of black holes of type D''},
\href{https://doi.org/10.1103/PhysRevD.104.084078}{Phys. Rev. D \textbf{104} (2021), 084078}
\href{https://arxiv.org/pdf/2108.02239}{\tt [arXiv:2108.02239 [gr-qc]]}.

\bibitem{Podolsky-2022}
J.~Podolsky and A.~Vratny,
{\it ``New form of all black holes of type D with a cosmological constant''}, 
\href{https://doi.org/10.1103/PhysRevD.107.084034}{Phys. Rev. D \textbf{107} (2023) no.8, 084034}
\href{https://arxiv.org/pdf/2212.08865}{\tt [arXiv:2212.08865 [gr-qc]]}.


\bibitem{bonnor}
W.~B.~Bonnor,
{\it ``A new interpretation of the NUT metric in general relativity''},
\href{https://doi.org/10.1017/s0305004100044807}{Math. Proc. Cambridge Phil. Soc. \textbf{66} (1969) no.1, 145-151}

\bibitem{belinski-book}
V. Belinski, E. Verdaguer, \href{https://doi.org/10.1017/CBO9780511535253}{\it ``Gravitational solitons''}, Cambridge, Cambridge Univ. Press, 2001.

\bibitem{marcoa-embedding}
M.~Astorino,
{\it ``Embedding hairy black holes in a magnetic universe''},
\href{https://doi.org/10.1103/PhysRevD.87.084029}{Phys. Rev. D \textbf{87} (2013) no.8, 084029} ;
\href{https://arxiv.org/pdf/1301.6794}{\tt [arXiv:1301.6794 [gr-qc]]}.

\bibitem{marcoa-stationary}
M.~Astorino,
{\it ``Stationary axisymmetric spacetimes with a conformally coupled scalar field''},
\href{http://doi.org/10.1103/PhysRevD.91.064066}{Phys. Rev. D \textbf{91} (2015), 064066} ;
\href{https://arxiv.org/pdf/1412.3539}{\tt [arXiv:1412.3539 [gr-qc]]}.

\bibitem{alekseev-belinski-kerr}
  G.~A.~Alekseev and V.~A.~Belinski,
  {\it ``Superposition of fields of two rotating charged masses in general relativity and existence of equilibrium configurations''},
  Gen.\ Rel.\ Grav.\  {\bf 51} (2019) no.5,  68
  \href{https://arxiv.org/abs/1905.05317}{\tt [arXiv:1905.05317 [gr-qc]]}.

\bibitem{bubble}
M.~Astorino, R.~Emparan and A.~Vigan\`o,
{\it ``Bubbles of nothing in binary black holes and black rings, and viceversa''},
\href{https://doi.org/10.1007/JHEP07(2022)007}{JHEP \textbf{07} (2022), 007} ;
\href{https://arxiv.org/pdf/2204.09690.pdf}{\tt [arXiv:2204.09690 [hep-th]]}.

\bibitem{tesi-giova}
G.~Boldi,
{\it ``Ehlers transformation and accelerating spacetimes with a gravomagnetic monopole''},
\href{https://inspirehep.net/literature/2652409}{\tt Università degli Studi di Milano (2022)}

\bibitem{charged-I}
M.~Astorino,
{\it ``Accelerating and Charged Type I Black Holes''},
\href{https://arxiv.org/pdf/2307.10534.pdf}{\tt[arXiv:2307.10534 [gr-qc]]}.

\bibitem{asto-lambda}
M.~Astorino,
{\it ``Charging axisymmetric space-times with cosmological constant''},
\href{https://doi.org/10.1007/JHEP06(2012)086}{JHEP \textbf{06} (2012), 086} ;
\href{https://arxiv.org/pdf/1205.6998.pdf}{\tt [arXiv:1205.6998 [gr-qc]]}.

\bibitem{adolfo}
J. Barrientos and A.~Cisterna,
{\it ``Ehlers Transformations as a Tool for Constructing Accelerating NUT Black Holes''},
\href{https://arxiv.org/pdf/2305.03765.pdf}{\tt[arXiv:2305.03765 [gr-qc]]}.

\bibitem{Griffiths:2009dfa}
J.~B.~Griffiths and J.~Podolsky,
{\it ``Exact Space-Times in Einstein's General Relativity''}, 
\href{https://doi.org/10.1017/CBO9780511635397}{Cambridge University Press, 2009}

\end{thebibliography}
\end{document}